\documentclass[11pt,a4paper]{article}
\usepackage{amsmath,amssymb,amsfonts,a4wide,graphicx,bm,times}
\usepackage{cite}

\makeatletter \let\old@startsection=\@startsection
\renewcommand{\@startsection}[6]
{\old@startsection{#1}{#2}{#3}{#4}{#5}{#6\mathversion{bold}}}
\makeatother

\makeatletter

\@addtoreset{equation}{section}
\makeatother
\let\refOld\ref
\renewcommand{\ref}[1]{(\refOld{#1})}

\newcommand{\superp}[2]{\genfrac{}{}{0pt}{}{#1}{#2}}

 \def\d{\delta}

 \def\Im{{\rm Im ~}}
 \def\p{\partial}
 
 \def\a{\alpha}
 \def\b{\beta}
 \def\g{\gamma}
 \def\d{\delta}
 \def\e{\epsilon}

 \def\l{\lambda}

 \def\s{\sigma}

 \def\G{\Gamma}
 \def\D{\Delta}

 \def\S{\Sigma}

\def\la{\left\langle}
\def\ra{\right\rangle}

\def\laN{\left\langle N\left|}
\def\raN{\right|N\right\rangle}
\def\hf{\dfrac{1}{2}}
\def\Op{\mathcal{O}}
\def\CD{\mathcal{D}}

\def\implies{\quad\Rightarrow\quad}

\def\vphi{\varphi}

\def\CF{\mathcal{F}}

\def\CZ{{\mathcal{Z}}}

\def\bens{$\b$-ensemble}

\def\pert{^{(\text{pert})}}

\def\inst{^{(\text{inst})}}
\def\full{^{(\text{full})}}

\def\brho{\bar\rho}

\def\bC{\bar C}

\def\CS{\mathcal{S}}

\def\SGC{\CS_\text{GC}}

\def\ZGC{\CZ_\text{GC}}
\def\ZC{\CZ_\text{C}}
\def\FGC{\CF_\text{GC}}
\def\FC{\CF_\text{C}}

\def\Li{\text{Li}_2}
\def\Qreg{Q_\text{reg.}}
\def\Qsing{Q_\text{sing.}}

\begin{document}
\begin{titlepage}
\renewcommand{\thefootnote}{\fnsymbol{footnote}}
\vspace*{-2cm}
\begin{flushright}
APCTP Pre2014-001
\end{flushright}

\vspace*{1cm}
    \begin{Large}
       \begin{center}
         {\huge Confinement and Mayer cluster expansions}
       \end{center}
    \end{Large}
\vspace{0.7cm}

\begin{center}
Jean-Emile B{\sc ourgine}\footnote
            {
e-mail address : 
jebourgine@apctp.org}\\
      
\vspace{0.7cm}                    
{\it Asia Pacific Center for Theoretical Physics (APCTP)
}\\
{\it Pohang, Gyeongbuk 790-784, Republic of Korea}
\end{center}

\vspace{0.7cm}

\begin{abstract}
\noindent
In these notes, we study a class of grand-canonical partition functions with a kernel depending on a small parameter $\e$. This class is directly relevant to Nekrasov partition functions of $\mathcal{N}=2$ SUSY gauge theories on the 4d $\Omega$-background, for which $\e$ is identified with one of the equivariant deformation parameter. In the Nekrasov-Shatashvili limit $\e\to0$, we show that the free energy is given by an on-shell effective action. The equations of motion take the form of a TBA equation. The free energy is identified with the Yang-Yang functional of the corresponding system of Bethe roots. We further study the associated canonical model that takes the form of a generalized matrix model. Confinement of the eigenvalues by the short-range potential is observed. In the limit where this confining potential becomes weak, the collective field theory formulation is recovered. Finally, we discuss the connection with the alternative expression of instanton partition functions as sums over Young tableaux.
\end{abstract}
\vfill

\end{titlepage}
\vfil\eject

\setcounter{footnote}{0}

\section{Introduction}
\begin{flushright}
\textit{``Lesson: If you are stupid, don't get discouraged but try a different way to get the result.''}\\
Madan Lal Mehta\footnote{Random Matrices, 3rd edition, appendix A.44} 
\end{flushright}

\hspace{2cm}

\noindent The recent developments in the study of $\mathcal{N}=2$ SUSY gauge theories in four dimensions have driven a lot of progress in several fields of theoretical physics. This is particularly the case for the field of random matrix models. The Alday-Gaiotto-Tachikawa (AGT) correspondence \cite{Alday2009,Wyllard2009,Alba2010,Fateev2011,Morozov2013} relates \bens\ partition functions to summations over Young tableaux of a deformed Plancherel measure \cite{Dijkgraaf2009}. This relation has been used extensively to verify the AGT proposal in various limits (see for instance \cite{Fujita2009,Mironov2010c,Mironov2010e,Itoyama2010a,Itoyama2011,Nishinaka2011,Bonelli2011,Bonelli2011a,Baek2013}). Similarly, the correspondence with quantum integrable systems proposed in \cite{Nekrasov2009} hints for a connection between the standard matrix models methods (such as loop equations \cite{Ginsparg1993} and collective field theory \cite{Jevicki1980,Jevicki1981}) and the Mayer cluster expansion \cite{Mayer1940,Mayer1941}. The latter is a statistical physics technique employed in \cite{Nekrasov2009} to treat the gauge theory partition functions. An investigation of this possible connection was performed in \cite{Bourgine2013}. There, the Mayer expansion of a grand-canonical (generalized) matrix model is compared to the canonical model at large $N$.\footnote{Here $N$ is the size of the matrix, it is the number of integration variables after diagonalization. It corresponds here to a number of instantons. It should not be mistaken with the rank $N_c$ of the gauge group, which will remain finite in this paper.} However, the results obtained in \cite{Bourgine2013} are not directly applicable to SUSY gauge theories. The goal of this paper is to extend some of these results to a larger class of models relevant to the gauge/integrability correspondence.

More precisely, the models considered here are defined as
\begin{equation}\label{def_CZ}
\ZGC(q)=\sum_{N=0}^\infty{\dfrac{q^N\e^{-N}}{N!}\ZC(N)},\quad \ZC(N)=\int_{\mathbb{R}^N}{\prod_{i=1}^NQ(\phi_i)\dfrac{d\phi_i}{2i\pi}\prod_{\superp{i,j=1}{i<j}}^N\left(1+\e f(\phi_i-\phi_j)\right)}.
\end{equation}
They will be studied in the limit $\e\to0$, with a $f(x)$ a sum of two terms:
\begin{equation}
f(x)=p(x)+G(x),\quad p(x)=\dfrac{\a\e}{x^2-\e^2}.
\end{equation}
The function $p(x)$ presents two single poles at $x=\pm\e$ with residue $\pm\a/2$. On the other hand, $G(x)$ is a symmetric function independent of $\e$, and such that $G(0)$ is finite. The parameter $\e$ is assumed to have a small positive imaginary part, and integrations in \ref{def_CZ} are contour integrals over the real axis. Following \cite{Moore1997}, contours are closed in the upper half-plane, and avoid possible singularities at infinity.\footnote{In particular, the volume integral is vanishing since the integrand does not have poles in the upper half plane, apart from the one at infinity,
\begin{equation}\label{zero_volume}
\int{\dfrac{d\phi}{2i\pi}}=0.
\end{equation}} Furthermore, both $Q(x)$ and $G(x)$ are supposed to have no singularities on $\mathbb{R}$. The presence of $p(x)$ was neglected in \cite{Bourgine2013}, and this particular case may be recovered by setting $\a=0$. In the limit $\e\to0$, the poles of $p(x)$ pinch the integration contour, drastically changing the behavior of the model. Physically, $p(x)$ gives rise to a strong integration at short distance $x\sim\e$, which can no longer be treated perturbatively as $\e p(x)$ becomes of order $O(1)$.

Instanton partition functions of $\mathcal{N}=2$ $SU(N_c)$ gauge theories reduces after localization to coupled one-dimensional integrals of the form \ref{def_CZ} with $\a=1$ \cite{Nekrasov2003}. This computation is regularized by considering the gauge theories on the $\Omega$-background which depends on two equivariant deformation parameters $\e_1,\e_2$. The Euclidean background $\mathbb{R}^4$ is recovered in the limit $\e_1,\e_2\to0$. Connections with quantum integrable systems appear when $\e_2$ is sent to zero while keeping $\e_1$ finite \cite{Nekrasov2009}. This limit is now referred as the Nekrasov-Shatashvili (NS) limit. Under the identification $\e=\e_2$, it coincides with the limit where the kernel becomes close to one considered here. The remaining deformation parameter $\e_1$ appears in the definition of the function $G(x)$. Its expression depends on the matter content of the gauge theory. It is given below in the limit $\e_2\to0$ for $\mathcal{N}=2$ super-Yang-Mills (SYM) with fundamental flavors, and $\mathcal{N}=2^\ast$ with an adjoint hypermultiplet of mass $m$,
\begin{align}\label{def_G}
\begin{split}
 &\mathcal{N}=2\ \text{SYM},\quad G(x)=\dfrac1{x+\e_1}-\dfrac1{x-\e_1},\\
 &\mathcal{N}=2^\ast,\quad G(x)=\dfrac{2\e_1 m(m+\e_1)(3x^2-m^2-\e_1m-\e_1^2)}{(x^2-\e_1^2)(x^2-m^2)(x^2-(m+\e_1)^2)}.
\end{split}
\end{align}
Let us also mention that setting $G=0$ while keeping $\a=1$, we recover the model proposed by J. Hoppe's in \cite{Hoppe1982}, and further studied in \cite{Kazakov1998,Hoppe1999}.

This paper is organized as follows. In the first part, we focus on the grand-canonical model and show that the free energy at first order in $\e$ takes the form of an on-shell effective action. For $\a=1$, we recover the action proposed in \cite{Nekrasov2009}, and the corresponding equations of motion take the form of a TBA-like relation. After a brief reminder on Bethe equations and Thermodynamical Bethe Ansatz (TBA), the identification between the free energy and the Yang-Yang functional is done. In the second section, we consider the associated canonical model. It is shown that the previous results can be derived from a confinement hypothesis for the eigenvalues of this model. We further study the limit $\a\to0$ and recover the collective field action of a Dyson gas \cite{Dyson1962}. Eventually, we compare our results to the alternative approach based on the expression of instanton partition functions as sum over Young tableaux. Appendices gather the most technical details.

\vspace{1cm}
\textit{As the present paper was in preparation, we received the preprint \cite{Meneghelli2013} where similar results are derived for Nekrasov partition functions. However, the method presented here is different from the one employed in \cite{Meneghelli2013}, and we believe it is interesting in its own right.}

\section{Grand canonical partition function in the NS limit}
The Mayer cluster expansion provides an expression of the grand-canonical partition function \ref{def_CZ} as a sum over clusters. These clusters consist of a set of vertices connected by at most one link. To each cluster is associated a set of coupled integrals, with a measure $Q(\phi_i)d\phi_i/2i\pi$ at each vertex $i$ and a kernel $\e f(\phi_i-\phi_j)$ for each link $<ij>$. Taking the logarithm reduces the summation to connected clusters, and the free energy writes \cite{Mayer1940,Mayer1941}
\begin{equation}\label{sum_clusters}
\FGC(q)=\e\log\ZGC(q)=\sum_{l=0}^\infty{q^l\sum_{\bC_l}\dfrac{\e^{-(l-1)}}{\s(\bC_l)}\int{\prod_{i\in V(\bC_l)}Q(\phi_i)\dfrac{d\phi_i}{2i\pi}\prod_{<ij>\in E(\bC_l)}\e f(\phi_i-\phi_j)}}.
\end{equation}
We denoted $\bC_l$ the connected clusters with $l$ vertices, and $V(\bC_l)$ (resp. $E(\bC_l)$) their set of vertices (resp. edges). The symmetry factor $\s(\bC_l)$ takes into account the possibilities of re-arranging the vertices, it is the cardinal of the group of automorphisms that preserve the cluster $\bC_l$. Since $f$ is a sum of two terms $G$ and $p$, the previous cluster expansion can be re-written as a sum over clusters $C_l$ with two types of links, referred as p- and G-links \cite{Andersen1977},
\begin{equation}\label{sum_clusters_II}
\FGC(q)=\sum_{l=0}^\infty{q^l\sum_{C_l}\dfrac{\e^{-(l-1)}}{\s(C_l)}\int{\prod_{i\in V(C_l)}Q(\phi_i)\dfrac{d\phi_i}{2i\pi}\prod_{<ij>\in E_p(C_l)}\dfrac{\a\e^2}{\phi_{ij}^2-\e^2}\prod_{<ij>\in E_G(C_l)}\e G(\phi_{ij})}},
\end{equation}
with the shortcut notation $\phi_{ij}=\phi_i-\phi_j$. Let us emphasize that clusters $C_l$ have the same structure as the previous clusters $\bC_l$, but with links bearing an additional label 'p' or 'G'. This labeling modify the symmetry factors, $\s(C_l)\leq\s(\bC_l)$ because automorphisms must preserve the edges labels. The set of edges $E(C_l)$ is obviously the direct union of the sets $E_p(C_l)$ and $E_G(C_l)$ of p- and G-links respectively.

\subsection{Structure of minimal clusters}
In order to compute $\FGC(q)$ at the leading order in $\e$, we need to determine the structure and order of clusters that give the minimal contribution for each term of the $q$-expansion. Such clusters will be referred shortly as 'minimal clusters'. Our strategy is to work recursively on the number of links. Connected clusters with the minimal number of links at fixed number of vertices have a tree structure. We will show that trees with $l$ vertices are of order $O(\e^{l-1})$, and then discuss the possibility to add either G- or p-links. Adding a link cannot decrease the order in $\e$, and we conclude that minimal clusters are of the order of the trees, i.e. $O(\e^{l-1})$. It implies that $\FGC(q)$ defined in \ref{sum_clusters} is of order one.

Clusters made of G-links were considered in \cite{Bourgine2013}. Each G-link brings a factor $\e$: G-trees are of order $\e^{l-1}$, and all G-clusters having cycles are sub-dominants. Trees made of p-links have integrals that are also of order $O(\e^{l-1})$. At this order, they can be computed recursively, this is done in appendix \refOld{AppA}. Their contribution only depends on the number $l$ of vertices, and read
\begin{equation}\label{p-trees}
\left(\dfrac{\a\e}{2}\right)^{l-1}\int{Q(\phi)^l\dfrac{d\phi}{2i\pi}}+O(\e^l).
\end{equation}
Trees involving both p- and G-links can be constructed recursively starting from a G-tree or a p-tree, and adding a number of vertices with G- or p-links. Adding G-links to a p-tree (or a mixed tree) will obviously bring a new factor $\e$ for each link. Adding p-links to G-trees (or mixed trees) is similar to adding p-links to p-trees (see section \refOld{A3} of the appendix), and also brings a factor $\e$ per link. Thus, any tree with $l$ vertices is of order $O(\e^{l-1})$.

\begin{figure}[!t]
\centering
\includegraphics[width=9cm]{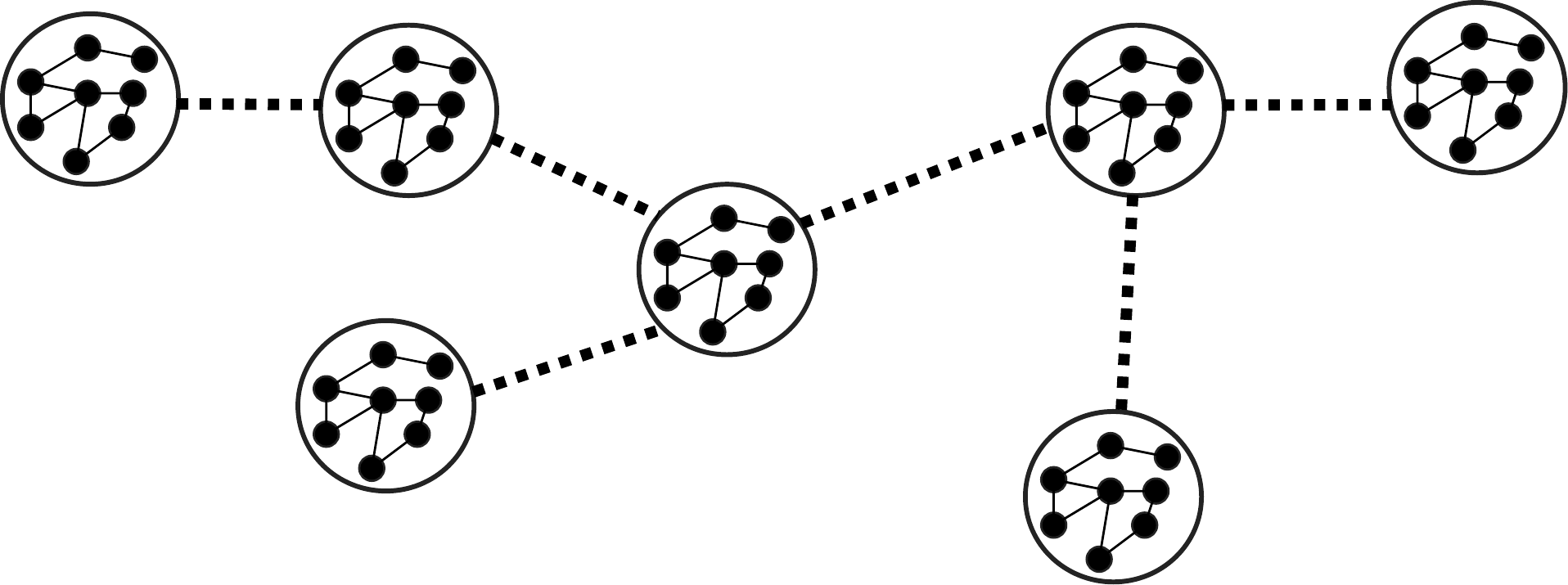}
\caption{General structure of minimal clusters. G-links are dashed and p-links in plain.}
\label{structure}
\end{figure}

Any cluster can be build from a tree by adding p- or G-links. Adding a G-link always cost a factor $\e$. It implies that $G$-links cannot belong to the cycles of minimal clusters. Otherwise they could be removed, leading to a connected cluster of strictly smaller order. Naively, adding a p-link brings a factor $O(\e^2)$. But kernel residues are proportional to $1/\e$, reducing the degree to $O(\e)$ which is still subleading. However, there is a specific case where adding a p-link is costless: when the vertices $x$ and $y$ between which the p-link is inserted are already connected by a path of p-links, thus forming a p-cycle. It is easy to convince oneself that this occurs by computing the integral associated to the first diagrams, such as the triangle, for a simple potential. In this case, the additional p-link is costless because if $x$ and $y$ are related through a path of p-links, taking the residues of the kernel poles along this path leads to the relation $y=x+k\e$, with $k$ an integer. The additional p-link is then evaluated at a fixed distance $k\e$ which is of order $\e p(k\e)=\a/(k^2-1)=O(1)$. Let us emphasize that this phenomenon cannot happen with G-links because $G(0)$ is finite. Note also that if the path between $x$ and $y$ had involved a G-link, the relation $y=x+k\e$ would no longer holds, and the cluster with additional p-link would be subleading. It is not possible to obtain a negative power of $\e$ by adding a p-link, and we conclude that minimal clusters are of order $O(\e^{l-1})$.\footnote{This will appear as a result of the next subsection, c.f. footnote \refOld{fnote}.}

To summarize, minimal clusters have the structure depicted in figure \refOld{structure}, with G-links connecting sets of vertices related by p-links. Cycles involve p-links but no G-links.

\subsection{Generating functions}
To derive the expression of the grand-canonical free energy at first order in $\e$, we employ the technique presented in \cite{Bourgine2013}. This method relies on generating functions of rooted clusters. Rooting a cluster $C_l$ into $C_l^x$ consists in marking the vertex $x$, such that it is left invariant under preserving automorphisms ($\s(C_l^x)\leq\s(C_l)$), and that its associated integration variable $\phi_x$ is fixed. With a slight abuse of notation, this integration variable will also be denoted $x$. The generating function of rooted clusters is defined as
\begin{equation}
Y(x)=q Q(x)\sum_{l=0}^\infty{\sum_{C_l^x}\dfrac{\e^{-(l-1)}}{\s(C_l^x)}\int{\prod_{i\in V(C_l^x)\smallsetminus\{x\}}qQ(\phi_i)\dfrac{d\phi_i}{2i\pi}\prod_{<ij>\in E_p(C_l^x)}\dfrac{\a\e^2}{\phi_{ij}^2-\e^2}\prod_{<ij>\in E_G(C_l^x)}\e G(\phi_{ij})}}.
\end{equation}
This expansion of $Y(x)$ as a sum over rooted clusters $C_l^x$ involves both p- and G-links. In order to treat separately the action of G and p kernels, it is useful to introduce two additional generating functions, denoted $qQ(x)Y_p(x)$ and $Y_G(x)$, by constraining the root $x$ to be tied to other vertices only through p-links and G-links respectively. All these generating functions are of order $O(1)$ in $\e$.

We have defined three functions $Y$, $Y_p$ and $Y_G$, and three relations among them will be derived: \ref{eom1}, \ref{eom_YG} and \ref{eom_Y}. In the next subsection, these functions will be used to express the free energy at small $\e$. The limit $\e\to0$ brings two main simplifications at the basis of our derivation: the tree structure for G-links, responsible for \ref{eom1} and \ref{eom_YG}, and the short range of p-interactions leading to \ref{eom_Y}.

\begin{figure}[!t]
\centering
\includegraphics[width=8cm]{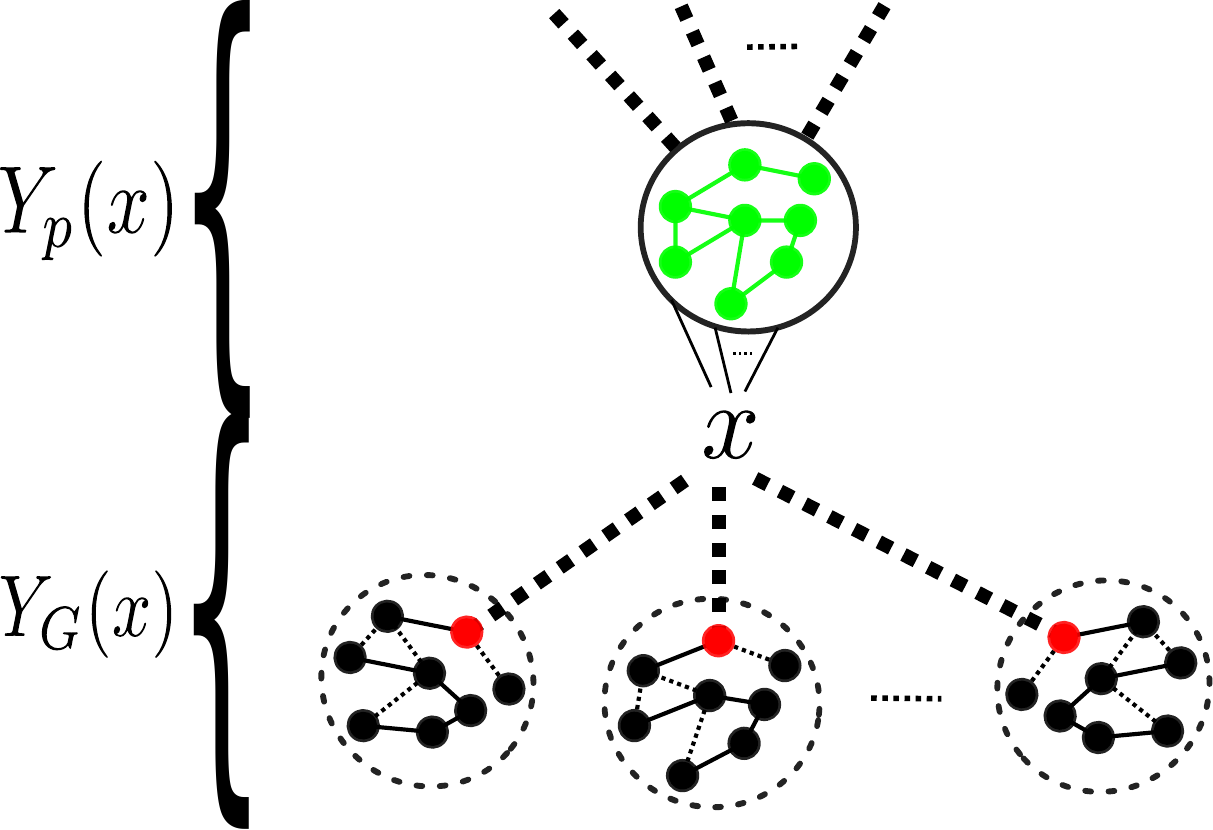}
\caption{A rooted minimal cluster. G-links are dashed and p-links in plain. Fixed vertices are in green, and direct G-vertices in red. Set of vertices containing only p-links are circled in plain, whereas set of vertices related through both p- and G- links have dashed circling.}
\label{rooted}
\end{figure}

Before giving the derivation of the three identities, we need to introduce some terminology. Vertices connected to the root by a single link are called \textit{direct vertices}. We further distinguish \textit{direct G-vertices} and \textit{direct p-vertices} by the type of link between the vertex and the root. Vertices related to the root through a path involving only p-links are called \textit{fixed vertices}. This set includes all the direct p-vertices. Finally, the \textit{descendants} of a vertex are the vertices connected to this vertex through a path that do not involve the root. The various types of vertices of a rooted clusters are highlighted in the figure \refOld{rooted}.

The simplest identity among $Y$, $Y_p$ and $Y_G$ is a factorization property at first order due to the fact that $G$-links cannot form cycles in minimal clusters,
\begin{equation}\label{eom1}
Y(x)\simeq Y_G(x)Y_p(x).
\end{equation}
Indeed, this identity expresses the fact that rooted clusters can be decomposed into two disconnected sub-cluster, as shown in figure \refOld{rooted}. The first sub-cluster involves the root, direct G-vertices and their descendant, whereas the second one contains the root, fixed vertices and their descendants. The identity \ref{eom1} will be used to eliminate $Y_p$, allowing to work only with $Y$ and $Y_G$.

The second identity is very similar to the one derived in \cite{Bourgine2013} for $\a=0$. It follows from the fact that the root in minimal clusters of $Y_G(x)$ can be connected to $m$ direct G-vertices, each being the root of a new cluster formed by their descendants. These new clusters are not connected to each other, otherwise we would form a cycle containing a G-link which would be subdominant. At first order, contributions of these clusters factorize, leading to
\begin{equation}\label{eom_YG}
Y_G(x)\simeq q Q(x)\sum_{m=0}^\infty{\dfrac{\e^{-m}}{m!}\prod_{i=1}^m\int{\e G(x-y_i)Y(y_i)\dfrac{dy_i}{2i\pi}}}=q Q(x)\exp\left(\int{G(x-y)Y(y)\dfrac{dy}{2i\pi}}\right).
\end{equation}
The symmetry factor $1/m!$ takes into account the possibility to permute the direct G-vertices.

To establish the last identity, we focus on the minimal clusters contributing to $Y_p(x)$. The set of fixed vertices form together with the root a rooted sub-cluster involving only $p$-links, that we denote $\D_l^x$ where $l$ is the number of fixed vertices plus the root $x$. Taking apart the p-links of $\D_l^x$, each fixed vertex becomes the root of a sub-cluster formed by itself, its direct G-vertices and their descendants. These sub-clusters are now disconnected, otherwise a cycle involving a G-link would appear. It implies again a factorization property at leading order,
\begin{equation}\label{Ap_exp}
Y_p(x)\simeq\sum_{l=1}^\infty\sum_{\D_l^x}\dfrac{\e^{-(l-1)}}{\s(\D_l^x)}\int{\prod_{i\in V(\D_l^x)\smallsetminus\{x\}}Y_G(\phi_i)\dfrac{d\phi_i}{2i\pi}\prod_{<ij>\in E(\D_l^x)}\dfrac{\a\e^2}{\phi_{ij}^2-\e^2}}.
\end{equation}
The next step is to argue that the factors $Y_G(\phi_i)$ within the integrals can be approximated by $Y_G(x)$ and taken out of the integrations. This seems reasonable because the kernel poles fix the relative value of the variables $\phi_i$ with respect to the root $x$ as $\phi_i=x+k_i\e$ with $k_i$ a finite integer. In the limit $\e\to0$, it is possible to approximate the residue factor $Y_G(x+k_i\e)$ with $Y_G(x)$. However, it turns out that this way of reasoning is too simple as it overlooks the contributions from the poles of $Y_G(\phi_i)$. These contributions are not negligible and even play a major role in the proper arguments given in appendix \refOld{AppA}. These arguments heavily rely on the hypothesis that the potential $Q(x)$ has only pole singularities within the contour of integration. It also requires that the kernel $G$ preserves this property under convolutions. Both requirements are satisfied in the application to instanton partition functions of $\mathcal{N}=2$ SUSY gauge theories. Replacing $Y_G(\phi_i)$ with $Y_G(x)$ in \ref{Ap_exp}, we obtain
\begin{equation}\label{factor_Yp}
Y_p(x)\simeq\sum_{l=1}^\infty c_l\e^{-(l-1)} Y_G(x)^{l-1},\quad c_l=\sum_{\D_l^x}\dfrac1{\s(\D_l^x)}\int{\prod_{i=1}^{l-1}\dfrac{d\phi_i}{2i\pi}\prod_{<ij>\in E(\D_l^x)}\dfrac{\a\e^2}{\phi_{ij}^2-\e^2}}.
\end{equation}
The coefficients $c_l$ will turn out to be independent of $x$. They can be related to the following integrals,
\begin{equation}
c_l=I_l\e^{l-1},\quad I_l=\dfrac1{(l-1)!\e^{l-1}}\int{\prod_{i=1}^{l-1}\dfrac{d\phi_i}{2i\pi}\prod_{\superp{i,j=1}{i<j}}^l\left(1+\dfrac{\a\e^2}{\phi_{ij}^2-\e^2}\right)},
\end{equation}
where we identified $\phi_l\equiv x$. To derive this relation, we apply the Mayer expansion to the integrals $I_l$. In the product, terms with $j=l$ act as a potential for the $l-1$ variables $\phi_{i<l}$. But we can take an alternate point of view and consider the expansion of the full product with $l$ variables over rooted clusters with root $\phi_l$. This expansion contains terms associated to disconnected clusters. Such terms factorize into connected part contributions. But contributions of clusters with no root vanishes because of the absence of poles in the potential.\footnote{These contributions read
\begin{equation}
\int{\prod_{i\in V(\D_k)}\dfrac{d\phi_i}{2i\pi}\prod_{<ij>\in E(\D_k)}\dfrac{\a\e^2}{\phi_{ij}^2-\e^2}}
\end{equation}
where $\D_k$ is a cluster of $k$ vertices and with only p-links. Poles in the kernel fixes the relative values of the variable $\phi_i$ in terms of one variable $\phi_1$, but there is no way to fix the remaining variable. This last integral is vanishing, due to the zero volume property \ref{zero_volume}.} Thus, only connected clusters remain, they can be identified to the clusters $\D_l^x$ considered above, and
\begin{equation}
I_l=\dfrac1{(l-1)!\e^{l-1}}\sum_{\D_l^x}n(\D_l^x)\int{\prod_{i\in V(\D_l^x)\smallsetminus\{x\}}\dfrac{d\phi_i}{2i\pi}\prod_{<ij>\in E(\D_l^x)}\dfrac{\a\e^2}{\phi_{ij}^2-\e^2}}.
\end{equation}
The integer $n(\D_l^x)$ is the number of configurations (i.e. labeled clusters) producing the same cluster $\D_l^x$. By definition, the symmetry factor $\s(\D_l^x)$ is the number of labellings of $l-1$ vertices (the root is fixed) divided by the number of equivalent configurations $n(\D_l^x)$, and thus $c_l=I_l\e^{l-1}$. The integral $I_l$ is a polynomial in $\a$ of degree $(l-1)(l-2)/2$ times $\a^{l-1}$, starting with the term\footnote{\label{fnote} Since $I_l=O(1)$, we have shown that minimal clusters with p-cycles are of order $O(\e^{l-1})$}
\begin{equation}\label{Il_1st}
I_l=\dfrac{1}{l!}\left(\dfrac{\a l}{2}\right)^{l-1}+O(\a^l).
\end{equation}
This expression is derived in appendix \refOld{AppB}. For the specific value $\a=1$ relevant to Nekrasov partition functions, we have the simple result $I_l=1/l$ also obtained in appendix \refOld{AppB}. For general $\a$, we introduce the function
\begin{equation}
l_\a(x)=\sum_{l=1}^\infty{I_l x^l},\quad l_1(x)=-\log(1-x).
\end{equation}
Inserting this expression in \ref{factor_Yp}, and using the factorization property \ref{eom1}, we conclude that $Y$ and $Y_G$ satisfy the identity
\begin{equation}\label{eom_Y}
Y(x)=l_\a(Y_G(x)).
\end{equation}
Together with \ref{eom_YG}, these two relations allow in principle to find the expression of the generating functions $Y(x)$ and $Y_G(x)$. At $\a=0$, there are no p-links, $Y_G(x)=Y(x)$, and we recover from \ref{eom_YG} the result of \cite{Bourgine2013}.

\subsection{Free energy}
\begin{figure}[!t]
\centering
\includegraphics[width=8cm]{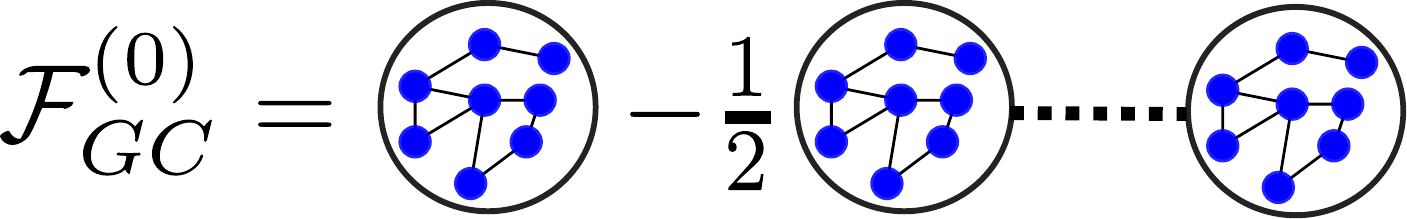}
\caption{The two contributions to the free energy at first order in $\e$. Blue vertices are dressed by G-links (in dashed), and related to each other with p-links (in plain). They form meta-vertices, drawn as black circles.}
\label{planar}
\end{figure}

It remains to relate the free energy to the generating functions $Y$ and $Y_G$. As can be seen in figure \refOld{structure}, the minimal clusters that contribute to the free energy at first order have a tree-structure in terms of G-links. We call meta-vertices the sets of vertices tied together by p-links. At large distance, figure \refOld{structure} resemble a tree with vertices replaced by meta-vertices. Thanks to this tree structure, the Basso-Sever-Vieira formula \cite{Basso2013} still applies. Indeed, the combinatorial argument given in the proof \cite{Bourgine2013} (appendix A) for the case $\a=0$ can be re-used for the general case since it is possible to associate uniquely a G-link to each meta-vertex. The inner structure of meta-vertices play no role in our discussion, and the symmetry factors are such that the free energy \ref{sum_clusters_II} is again given by a difference of two terms,
\begin{equation}
\FGC^{(0)}=\lim_{\e\to0}\FGC=\G_0-\hf\G_1,
\end{equation}
This equation is graphically represented in figure \refOld{planar}. In the RHS, the second term $\G_1$ is obtained by attaching two rooted clusters through a G-link. Since the roots belong to two different meta-vertices,
\begin{equation}
\G_1=\int{Y(x)Y(y)G(x-y)\dfrac{dxdy}{(2i\pi)^2}}.
\end{equation}
To obtain the term $\G_0$, focus on a meta-vertex. Taking apart its p-links, the inner vertices (in blue on figure \refOld{planar}) are the roots of disconnected sub-clusters with direct vertices of type G only. It means that they should be dressed with $Y_G(x)$, and the first term is
\begin{equation}
\G_0=\sum_{l=1}^\infty\sum_{\D_l}\dfrac{\e^{-(l-1)}}{\s(\D_l)}\int{\prod_{i\in V(\D_l)} Y_G(\phi_i)\dfrac{d\phi_i}{2i\pi}\prod_{<ij>\in E(\D_l)}\dfrac{\a\e^2}{\phi_{ij}^2-\e^2}},
\end{equation}
where $\D_l$ is the connected cluster with only p-links associated to a meta-vertex. One possibility to compute this expression is to again approximate $Y_G(\phi_i)\simeq Y_G(x)$ and evaluate the remaining integrals, but the justifications may be tedious. Fortunately, it is possible to make use of a result demonstrated in \cite{Bourgine2013} (appendix B) which relates the symmetry factors of rooted clusters with non-rooted ones,
\begin{equation}
\dfrac{l}{\s(\D_l)}=\sum_{k}\dfrac1{\s(\D_l^{x_k})}.
\end{equation}
In this formula, the summation is over the class of equivalence of rooted clusters obtained by rooting a vertex $x_k$ of the original cluster $\D_l$. This identity allows to single out one of the integration variable, and rewrite $\G_0$ as
\begin{equation}
\G_0=\int{\dfrac{dx}{2i\pi}Y_G(x)\sum_{l=1}^\infty\dfrac1l\sum_{\D_l^x}\dfrac{\e^{-(l-1)}}{\s(\D_l^x)}\int{\prod_{i\in V(\D_l^x)\smallsetminus\{x\}} Y_G(\phi_i)\dfrac{d\phi_i}{2i\pi}\prod_{<ij>\in E(\D_l^x)}\dfrac{\a\e^2}{\phi_{ij}^2-\e^2}}}.
\end{equation}
We recognize the integrals that appeared in the expression \ref{Ap_exp} of $Y_p(x)$. Using the same trick to evaluate them, we get
\begin{equation}
\G_0=\int{\dfrac{dx}{2i\pi}L_\a(Y_G(x))},
\end{equation}
where we introduced the primitive
\begin{equation}\label{def_La}
L_\a(x)=\sum_{l=1}^\infty{I_l\dfrac{x^l}{l}}\implies x\p_xL_\a(x)=l_\a(x).
\end{equation}
At $\a=1$, $I_l=1/l$ and $L_\a$ is equal to the dilogarithm function $\Li$. The additional factor $1/l$ necessary to transform $l_\a$ into $L_\a$, and thus generate the $\Li$ at $\a=1$, comes from the choice of a root among the $l$ inner vertices of a meta-vertex.

The expression of the grand-canonical free energy takes a nice form if we re-introduce the relation \ref{eom_YG} between $Y$ and $Y_G$, to write it as
\begin{equation}\label{FE}
\FGC^{(0)}=\hf\int{Y(x)Y(y)G(x-y)\dfrac{dxdy}{(2i\pi)^2}}+\int{Y(x)\log(qQ(x))\dfrac{dx}{2i\pi}}+\int{\left[L_\a(Y_G(x))-Y(x)\log Y_G(x)\right]\dfrac{dx}{2i\pi}}.
\end{equation}
Instead of $Y$ and $Y_G$, let us define $\rho$ and $\vphi$ as
\begin{equation}\label{def_rp}
Y(x)=2i\pi\rho(x),\quad Y_G(x)=qQ(x)e^{-\vphi(x)}.
\end{equation}
The function $\rho(x)$ is interpreted as the grand-canonical instanton density (see appendix \refOld{AppC} or \cite{Bourgine2013}). This change of variables allows to re-write $\FGC^{(0)}$ as an on-shell action, i.e.
\begin{equation}
\FGC^{(0)}=\SGC[\rho^\ast,\vphi^\ast],\quad\text{with}\quad \left.\dfrac{\d\SGC}{\d\rho}\right|_{\rho=\rho^\ast}=0,\quad,\left.\dfrac{\d\SGC}{\d\vphi}\right|_{\vphi=\vphi^\ast}=0,
\end{equation}
and
\begin{equation}\label{action}
\SGC[\rho,\vphi]=\hf\int{\rho(x)\rho(y)G(x-y)dxdy}+\int{\rho(x)\vphi(x)dx}+\int{L_\a(qQ(x)e^{-\vphi(x)})\dfrac{dx}{2i\pi}}.
\end{equation}
The equations of motion reproduce the two relations \ref{eom_Y} and \ref{eom_YG} obtained previously. At the special value $\a=1$, we recover the results presented in \cite{Nekrasov2009} (formula 6.1).

\subsection{Relation with the TBA}\label{sec_Bethe}
At $\a=1$, the equations of motion \ref{eom_YG} and \ref{eom_Y} imply that the function $\vphi(x)$ defined in \ref{def_rp} obeys a Non-Linear Integral Equation (NLIE),
\begin{equation}\label{NLIE}
\vphi(x)=\int{G(x-y)\log\left(1-qQ(y)e^{-\vphi(y)}\right)\dfrac{dy}{2i\pi}}
\end{equation}
This equation appears in the Thermodynamical Bethe Ansatz method developed in \cite{Yang1968,Zamolodchikov1990a}, usually written in terms of the pseudo-energy $\e(x)=\vphi(x)-\log(qQ(x))$ with rapidity $x$. It does not pertain to a specific integrable model, but, on the opposite, provides a way to unify the description of different models \cite{Destri1994}. It appears here with an arbitrary potential $Q(x)$ and kernel $G(x)$, provided they satisfy the proper analyticity conditions necessary to derive \ref{eom_Y}. In this subsection, we recall how the NLIE can be obtained from a general set of Bethe equations. From the comparison with our previous results, a dictionary will be established with several integrable model quantities. In particular, we will identify the free energy $\FGC^{(0)}$ with the Yang-Yang functional introduced in \cite{Yang1968}. Although these results are not new and can be found in the existing literature on integrability, we present them here for completeness.\footnote{We would like to thank Dima Volin for his kind and patient explanations.}

We restrict ourselves to the kernel given in \ref{def_G} for $\mathcal{N}=2$ SYM. Consider the variables $u_i$, $i=1\cdots M$, satisfying a set of $M$ equations,
\begin{equation}\label{Bethe_equ}
1=qQ(u_i)\prod_{j=1}^M\dfrac{u_i-u_j-\e_1}{u_i-u_j+\e_1}.
\end{equation}
By analogy with integrable systems, the variables $u_i$ will be called \textit{Bethe roots} and the equations \ref{Bethe_equ} \textit{Bethe equations}. Depending on the form of the potential $Q(x)$, and the value of $\e$, these equations may be relevant to the system of bosons in 1d with $\d$-interaction (or quantum non-linear Schr\"odinger equation), or to the XXX (or $s\ell_2$) spin chain \cite{Gaudin1983}. The parameter $q$ is interpreted as a twist of the periodic boundary conditions. To establish the NLIE associated to \ref{Bethe_equ}, we employ a trick that goes back to \cite{Destri1994}.\footnote{This method was then generalized in \cite{Fioravanti1996} by including the presence of holes to treat excited states, and later on in \cite{Bombardelli2008} in the context of $\mathcal{N}=4$ SYM. For a nice review, see \cite{Fioravanti2010}.} First, by taking the logarithm, we find
\begin{equation}\label{log_Bethe}
2i\pi\eta_i=\log qQ(u_i)+\sum_{j=1}^M\log\left(\dfrac{u_i-u_j-\e_1}{u_i-u_j+\e_1}\right),
\end{equation}
where $\eta_i$ is an integer. It leads to define the counting function
\begin{equation}\label{def_eta}
2i\pi\eta(x)=\log qQ(x)+\log\left(\dfrac{q(x-\e_1)}{q(x+\e_1)}\right),\quad q(x)=\prod_{i=1}^M(x-u_i),
\end{equation}
where $q(x)$ is the Baxter Q-function, a monic polynomial with zeros at the Bethe roots position. From \ref{log_Bethe}, we deduce that $\eta(x)$ is an integer $\eta_i$ at $x=u_i$. But it may also be an integer for other values of $x$, and we should introduce a contour $\G$ that surrounds only the set of Bethe roots. The function
\begin{equation}\label{func_eta}
\dfrac1{1-e^{-2i\pi \eta(x)}}
\end{equation}
has poles for $\eta(x)$ integer, with residue $1/2i\pi\eta'(x)$ so that for any function $r(x)$ without singularity in the domain delimited by the contour $\G$, we have
\begin{equation}
\sum_{i=1}^Mr(u_i)=\oint_\G{\dfrac{r(x)}{1-e^{-2i\pi\eta(x)}}\eta'(x)dx}.
\end{equation}
Applying this result to \ref{def_eta}, we get after integration by parts the integral equation
\begin{equation}\label{TBA}
2i\pi\eta(x)=\log qQ(x)-\oint_\G{G(x-y)\log\left(1-e^{2i\pi\eta(y)}\right)dy},
\end{equation}
with $G(x)$ given in the first line of \ref{def_G}. It must emphasized that this NLIE has been obtained without taking the thermodynamical limit $M\to\infty$. In order to compare with the gauge theory result in \ref{NLIE}, assumptions of continuity for $\eta(x)$ are superfluous, and it is perfectly safe to work with a discrete set of roots. Actually, the condensations of roots only appear when we further send $\e_1\to0$ \cite{Bourgine2012a}.\footnote{See also the subsection \refOld{sec_dens} below for the identification of densities.} Comparing \ref{NLIE} with \ref{TBA} leads to identify:
\begin{equation}\label{ident}
2i\pi\eta(x)=-\e(x)=\log qQ(x)-\vphi(x),\quad 2i\pi\rho(x)=-\log\left(1-e^{-\e(x)}\right).
\end{equation}
In addition, the contours of integration must coincide, implying that all the Bethe roots satisfy $0<\Im u_i<\infty$, and that they are the only singularities of \ref{func_eta} in the upper half-plane. Then \ref{NLIE} corresponds to the NLIE of TBA in the bosonic case, for a scattering amplitude $S(x)$ such that $G(x)=\p\log S(x)$ \cite{Zamolodchikov1990a}. The even parity of $G(x)$ is a consequence of the unitarity of $S(x)$. We may further observe that the density of Bethe roots is (minus) the derivative of $\rho(x)$, and the logarithm of the Baxter q-function is the resolvent associated to $\rho(x)$,
\begin{equation}\label{rel_densities}
\rho_B(x)=\sum_{i=1}^M{\d(x-u_i)}=-\dfrac{d}{dx}\rho(x),\quad \log q(x)=-\int{\dfrac{\rho(y)}{x-y}dy}.
\end{equation}

The relations found previously between the gauge theory densities and the Bethe roots system allow to interpret the free energy \ref{FE} as a Yang-Yang functional. Indeed, from the identification \ref{ident}, we find after integration by parts,
\begin{align}
\begin{split}
&\int{L_\a(Y_G(x))\dfrac{dx}{2i\pi}}=2i\pi\int{\eta(x)\rho(x)dx}+2i\pi\int{x\eta(x)\rho'(x)dx},\\
&\int{Y(x)\log Y_G(x)\dfrac{dx}{2i\pi}}=2i\pi\int{\eta(x)\rho(x)dx}.
\end{split}
\end{align}
Taking the difference, and using the relation \ref{rel_densities} with the Bethe roots density leads to
\begin{equation}
\int{\left[L_\a(Y_G(x))-Y(x)\log Y_G(x)\right]\dfrac{dx}{2i\pi}}=-2i\pi\sum_{i=1}^M{\eta_iu_i}.
\end{equation}
Coming back to the expression \ref{FE} of the free energy, the Bethe roots density can be introduced using again the integration by parts, and 
\begin{equation}\label{YY}
\FGC^{(0)}=-\hf\sum_{i,j=1}^MG_{II}(u_i-u_j)+\sum_{i=1}^MV_I(u_i)-2i\pi\sum_{i=1}^M{\eta_iu_i}.
\end{equation}
where we used the primitives
\begin{equation}
\p_xV_I(x)=\log qQ(x),\quad \p_x^2G_{II}(x)=G(x).
\end{equation}
This is indeed the Yang-Yang functional first defined in \cite{Yang1968}.\footnote{Let us illustrate this in the case of bosons with $\d$ interaction. For simplicity, we consider an odd number of bosons, and an even number of Bethe roots, satisfying the equations
\begin{equation}
e^{iLu_i}=\prod_j\dfrac{u_i-u_j+ic}{u_i-u_j-ic}.
\end{equation}
It corresponds to \ref{Bethe_equ} with $\e_1=ic$ and $\log qQ(x)=ixL$ where $L$ is the volume. Up to a factor $i$, \ref{YY} is equivalent to the Yang-Yang functional:
\begin{equation}
-i\FGC^{(0)}=\dfrac{L}{2}\sum_{i=1}^M{u_i^2}-2i\pi\sum_{i=1}^M\eta_iu_i+\sum_{i,j=1}^M\int_0^{u_i-u_j}\tan^{-1}(k/c)dk.
\end{equation}}

\section{Canonical partition function}
In this section, we turn to the canonical ensemble and study the partition function $\ZC(N)$ defined in \ref{def_CZ}. The results obtained previously for the free energy at small $\e$ will be recovered from the expression of $\ZC(N)$ using a simple confinement hypothesis. This phenomenon of confinement was called \textit{instanton clustering} in \cite{Nekrasov2009}. \footnote{Due to the abundance of the word \textit{cluster} in this paper, we prefer to use the term \textit{confinement} instead of \textit{clustering} as both are appropriate.} It is the underlying cornerstone of the derivation presented in the previous section, it notably appeared in the step leading to \ref{factor_Yp}. The use of generating functions in this derivation brought a stronger justification for the validity of this hypothesis.

The study of the canonical partition function also provides a nice interpretation of the fields $\rho$ and $\vphi$ in the action \ref{action}. Furthermore, this action reduces to the collective field action of the canonical partition function $\ZC(N)$ in the appropriate limit. This is an interesting check of the results obtained before. Finally, we shall comment on an alternative approach \cite{Poghossian2010,Fucito2011} based on the evaluation of the canonical partition function as a sum over residues that are in one-to-one correspondence with boxes of a set of Young tableaux. 

\subsection{Confinement}
To study the canonical partition function, we start from the results obtained on the grand-canonical side, and invert the discrete Laplace transform with the formula
\begin{equation}\label{invert_Laplace}
\ZC(N)=N!\e^{N}\oint_0{\dfrac{dq}{2i\pi q}q^{-N}\ZGC(q)}.
\end{equation}
Since confinement originates from the p-term of the kernel, it is better understood when the G-interaction is turned off. We will first examine this simpler case, and later re-introduce the G-term. Setting bluntly $G=0$ in \ref{action}, the equations of motion imply $\vphi(x)=0$ and only remains
\begin{equation}
\FGC^{(0)}=\int{L_\a(qQ(x))\dfrac{dx}{2i\pi}}.
\end{equation}
The definition \ref{def_La} of the function $L_\a$ provides the $q$-expansion of the free energy. After exponentiation, the $q$-expansion of the grand-canonical partition function can be plugged into the inversion formula \ref{invert_Laplace} to give
\begin{equation}\label{exp_ZC_noG}
\ZC(N)=\sum_{p=1}^N\dfrac1{p!}\sum_{\superp{k_1,k_2,\cdots, k_p=1}{\sum k_i=N}}^N\dfrac{N!}{\prod_i k_i!}\int{\prod_{i=1}^pJ_{k_i}(x_i)Q(x_i)^{k_i}\dfrac{dx_i}{2i\pi}},
\end{equation}
where we denoted $J_k(x)$ the following integral,
\begin{equation}
J_k(x)=\int{\prod_{i=1}^{k-1}\dfrac{d\phi_i}{2i\pi}\prod_{\superp{i,j=1}{i<j}}^k(1+\e p(\phi_{ij}))},\quad\text{with}\quad \phi_k\equiv x.
\end{equation}
This integral equals to $(k-1)!\e^{k-1}I_k$ which is actually independent of $x$.

We would like to compare the expression \ref{exp_ZC_noG} with the original definition \ref{def_CZ} of $\ZC(N)$ for $G=0$. In the Dyson gas interpretation \cite{Dyson1962}, \ref{def_CZ} describes a gas of particles in one dimension, with position $\phi_i$, in an external potential $\log Q(\phi_i)$, and interacting through the kernel $p$. For standard matrix models, these particles are associated to the eigenvalues of the matrix. Here, it is more suitable to call them \textit{quarks}, by (rough) analogy with the confinement in QCD. Contrary to the G-interaction which remains weak for any distance $|\phi_i-\phi_j|$, the p-interaction becomes strong at small distance, i.e. when $|\phi_i-\phi_j|\propto\e$. Thus, the particles are expected to be confined, forming hadrons of an arbitrary number of quarks. Since hadrons are supposed to be of size $\sim\e$, quarks in the same hadrons experience an equal potential (at first order in $\e$). Then, the total potential associated to a hadron of $k$ quarks at position $x$ is approximately $J_k(x)Q(x)^k$. The factor $Q(x)^k$ is simply the product of the potentials that each quark feels, taken at the center of mass $x$. The additional factor $J_k(x)$ reflects the inner structure of hadrons, it takes into account the interactions between quarks through p-links.

Assume that $p$ hadrons are formed, each containing $k_i$ quarks. Using the confinement approximation for \ref{def_CZ}, we find the integral in \ref{exp_ZC_noG}. In this approximation, the p-interaction between hadrons is neglected because $\e p(\phi_{ij})$ is of order $O(\e^2)$ for a finite distance $|\phi_i-\phi_j|$. To fully obtain \ref{exp_ZC_noG}, it only remains to multiply by the appropriate combinatorial coefficients ($p!$ and $k_i!$ for respectively the indistinguishability of hadrons and quarks within a hadron), and sum over the possible configurations. This shows that the expression \ref{exp_ZC_noG}, and by extension the action \ref{action} for the grand-canonical ensemble, can be derived from the definition \ref{def_CZ} using the confinement hypothesis.

The same conclusions are reached when the G-interaction is turned back on. At first order in $\e$, the grand-canonical partition function $\ZGC(q)$ can be expressed as a path integral over the field $\rho$ and $\vphi$, with the action \ref{action},
\begin{equation}\label{ZGC_coll}
\ZGC(q)\simeq\int{D[\rho,\phi]\exp\dfrac1\e\left(\hf\int{\rho(x)G(x-y)\rho(y)dxdy}+\int{\rho(x)\vphi(x)dx}+\int{L_\a(qQ(x)e^{-\vphi(x)})\dfrac{dx}{2i\pi}}\right)}.
\end{equation}
Expanding the exponential of the function $L_\a$, but keeping the other terms, we find
\begin{equation}
\ZGC(q)\simeq\int{D[\rho,\vphi]e^{\frac1{2\e}\rho G\rho+\frac1\e \rho\vphi}\left(1+\sum_{p=1}^\infty{\dfrac{\e^{-p}}{p!}\sum_{\{ k_i\}}\prod_{i=1}^p\dfrac{I_{k_i}}{k_i}\int{\left(qQ(x_i)e^{-\vphi(x_i)}\right)^{k_i}\dfrac{dx_i}{2i\pi}}}\right)},
\end{equation}
where we used the simplified notations $\rho G\rho$ for the G-kernel (first term in the exponential \ref{ZGC_coll}) and $\rho\vphi$ for the source term (second term in \ref{ZGC_coll}). Using again the inversion formula \ref{invert_Laplace}, we deduce the expression of the canonical partition function,
\begin{equation}\label{ZC_G}
\ZC(N)\simeq\sum_{p=1}^N\dfrac1{p!}\sum_{\superp{k_1,k_2,\cdots, k_p=1}{\sum k_i=N}}^N\dfrac{N!}{\prod_i k_i!}\int{\prod_{i=1}^pJ_{k_i}(x_i)Q(x_i)^{k_i}\dfrac{dx_i}{2i\pi}\int{D[\rho,\vphi]e^{\frac1{2\e}\rho G\rho}e^{\frac1\e\int{\vphi(x)\left[\rho(x)-\e\sum_ik_i\d(x-x_i)\right]dx}}}}.
\end{equation}
The field $\vphi(x)$ appears to be a Lagrange multiplier enforcing the equality
\begin{equation}\label{rho_k}
\rho(x)=\e\sum_{i=1}^p k_i\d(x-x_i).
\end{equation}
The function $\rho(x)$ is interpreted as a density of quarks where the positions of confined quarks are replaced by their center of mass,
\begin{equation}
\rho(x)=\e\sum_{\a=1}^N\d(x-\phi_i)\simeq \e\sum_{i=1}^pk_i\d(x-x_i).
\end{equation}
This interpretation is not a surprise, given the definition \ref{def_rp} and the results of \cite{Bourgine2013}, briefly summarized in appendix \refOld{AppC}: $\rho(x)$ coincides with the grand-canonical density $\rho_{GC}(x)$ defined in \ref{def_rGC} as the vev of the quarks density operator. At first order, this density can be replaced by a density of hadrons weighted by their number of quarks. This density enters in the expression of the kernel term $\rho G\rho$, implying that G-interactions of quarks can be approximated by an interaction between hadrons. Indeed, plugging the expression \ref{rho_k} of $\rho(x)$ into \ref{ZC_G} gives
\begin{equation}
\ZC(N)=\sum_{p=1}^N\dfrac1{p!}\sum_{\superp{k_1,k_2,\cdots, k_p=1}{\sum k_i=N}}^N\dfrac{N!}{\prod_i k_i!}\int{\prod_{i=1}^pJ_{k_i}(x_i)Q(x_i)^{k_i}\dfrac{dx_i}{2i\pi}\exp\left(\dfrac{\e}{2}\sum_{i,j}{k_ik_j G(x_i-x_j)}\right)}.
\end{equation}
This expression can be obtained from the definition \ref{def_CZ} using again the confinement hypothesis.\footnote{To treat the kernel, we also need to assume
\begin{equation}
(1+\e G+\e p)\simeq(1+\e p)(1+\e G),
\end{equation}
i.e. to neglect the term $\e^2 p G$. It is possible because this term is a multiplicative correction of order $O(\e)$ to $\e p$ which is of order one only within cycles where $O(\e)$ corrections are subleading.}

\subsection{Collective field theory and $\a\to0$ limit}
It is natural to wonder how the results of \cite{Bourgine2013} concerning the comparison with matrix model techniques extend to the present model. Here we briefly discuss the collective field theory of the canonical model \cite{Jevicki1981}. The collective action presented in \cite{Bourgine2013} is not adapted to the treatment of the model \ref{def_CZ}. Indeed, it only reproduces the sum over clusters with tree structures in the Mayer expansion, thus overlooking the cycles involving p-links. Furthermore, it assumes that $\e p$ can be treated perturbatively, so that $1+\e p\simeq e^{\e p}$ at first order, which is not valid within a p-cycle where $\e p=O(1)$. In order to retrieve the collective field theory of \cite{Bourgine2013}, we need to impose $\e p\ll1$ for any link, which can be achieved by sending $\a\to0$. To have the G-links of same order, we rescale $G\to\a G$. It is also necessary to renormalize the fugacity $q\to q/\a$, the free energy $\FGC\to\a\FGC$ and the density $\rho\to\a\rho$.  As a result, the grand-canonical free energy at first order in $\a$ (and $\e$) is given by a summation over clusters with a tree structure, involving both p- and G-links, which can be compared with the collective field theory of the canonical model.

Using the asymptotic \ref{Il_1st} for $I_l$, the function $l_\a$ as $\a\to0$ can be approximated by a tree function,
\begin{equation}
l_\a(x/\a)\simeq \dfrac2\a T(x/2).
\end{equation}
The tree function \cite{Corless1997} is related to the principle branch of the Lambert W function through $T(x)=-W(-x)$. It satisfies the following properties,
\begin{equation}
T(x)=\sum_{n=1}^\infty\dfrac{x^n n^n}{n\times n!},\quad T(x)e^{-T(x)}=x,\quad \sum_{n=1}^\infty\dfrac{x^n n^n}{n^2\times n!}=T(x)\left(1-\hf T(x)\right),
\end{equation}
that can be used, together with the identity \ref{eom_YG}, to eliminate $Y_G$ from the expressions
\begin{equation}
L_\a(Y_G(x))=Y(x)\left(1-\dfrac\a4Y(x)\right),\quad \log Y_G(x)=\log\left(\a Y(x)\right)-\dfrac\a2Y(x).
\end{equation}
The free energy \ref{FE} simplifies as $\a\to0$ into
\begin{equation}\label{FGC_a0}
\FGC^{(0)}=\hf\int{\rho(x)\rho(y)G(x-y)dxdy}+\int{\rho(x)\log\left(\dfrac{qQ(x)}{2i\pi}\right)dx}+\dfrac{i\pi}2\int{\rho(x)^2dx}-\int{\rho(x)\left[\log\rho(x)-1\right]dx},
\end{equation}
where we used \ref{def_rp} to replace $Y$ with $\rho$. This expression should be compared with the collective action associated to the partition function
\begin{equation}\label{ZC_order1}
\ZC(N)\simeq\int{\prod_{i=1}^N Q(\phi_i)\dfrac{d\phi_i}{2i\pi}\prod_{\superp{i,j=1}{i<j}}^Ne^{\e\a G(\phi_{ij})+\e p(\phi_{ij})}}.
\end{equation}
We analyze the expression \ref{FGC_a0} term by term. The first term is obviously associated to the double sum of $G(\phi_{ij})$ in the exponential. Remember that the grand-canonical density $\rho(x)$ is equal to the canonical density defined in \ref{rel_rho}, up to a normalization factor $\g$ which also appears in the relation \ref{rel_FE} between free energies. The second term in \ref{FGC_a0} is a potential term generated by the product of $Q(\phi_i)$ in \ref{ZC_order1}. The third term comes from the double sum of $p(\phi_{ij})$ using the regularization $p(x)\simeq i\pi\a\d(x)$. This approximation cannot be made in the general model \ref{def_CZ}, it is only valid for p-links in tree structures. Finally, the last term, also called \textit{entropic term}, is a Gibbs factor. It can be computed from the change of integration measure $\prod_id\phi_i\to D[\rho]$ (see \cite{Bourgine2013}, appendix C for a derivation). We have thus verified that \ref{FGC_a0} coincide, in the sense of \cite{Bourgine2013}, with the collective field action of the canonical model.

\subsection{Parallel with the sum over Young tableaux}\label{sec_dens}
The contour integrals involved in the expression of the Nekrasov instanton partition functions can be evaluated exactly. The residues are in one-to-one correspondence with the boxes of a set of Young tableaux \cite{Nekrasov2003}. This expression of the canonical partition function is at the origin of an alternative approach to the NS limit. This approach was first employed by Nekrasov and Okounkov in \cite{Nekrasov2003a} to recover the Seiberg-Witten prepotential \cite{Seiberg1994,Seiberg1994a} from the instanton partition function in the $\mathbb{R}^4$ limit $\e_1,\e_2\to0$ of the $\Omega$-background. It was then extended to the NS limit $\e_2\to0$ in \cite{Poghossian2010,Fucito2011,Bourgine2012a,Ferrari2012a}.\footnote{There exists yet another approach for which the sums over Young tableaux are transformed back into matrix model integrals \cite{Klemm2008,Sulkowski2009,Sulkowski2009a}.} For simplicity, here we restrict ourselves to $\mathcal{N}=2$ $SU(N_c)$ SYM with $N_f$ fundamental flavors. However, the method is much more general, and applies to quiver theories as well \cite{Nekrasov2012,Fucito2012,Nekrasov2013}. In this approach, the free energy is given in terms of a function satisfying a Baxter TQ relation. Here, we explain how this equation relates to the system of Bethe roots considered in the subsection \refOld{sec_Bethe}. The purpose of this subsection is not to provide a rigorous derivation, but simply to illustrate the consequences of a relation between densities. Accordingly, we will skip the treatment of the perturbative part of the gauge partition function, and only refer to \cite{Bourgine2012a} for more details.

Super Yang-Mills with gauge group $SU(N_c)$ and $N_f$ hypermultiplets in the fundamental representation is sometimes called super-QCD. Its potential is a ratio of mass and gauge polynomials,
\begin{equation}
Q(x)=\dfrac{\prod_{f=1}^{N_f}(x-m_f)}{A(x+\e_1+\e_2)A(x)},\quad A(x)=\prod_{l=1}^{N_c}(x-a_l),
\end{equation}
where $m_f$ denotes the hypermultiplets masses, and $a_l$ the Coulomb branch vevs. Singularities on the real line are moved away by a small shift $a_l\to a_l+i0$ such that poles at $x=a_l$ are inside the integration contour, but not those at $x=a_l-\e_1-\e_2$. The G-kernel in the limit $\e_2\to0$ is given by the first line of \ref{def_G}, and $\a=1$. The residues $\phi_I$ are labeled by the multiple index $(l,i,j)$ where $l=1\cdots N_c$ is a color index, and $(i,j)$ denotes a box in the $l$th Young tableaux $\l^{(l)}$. They are interpreted as the instantons position in the moduli space, and are given by
\begin{equation}
\phi_{l,i,j}=a_l+(i-1)\e_1+(j-1)\e_2.
\end{equation}
We will further denote $\l_i^{(l)}$, $\l_i^{(l)}\geq\l_{i+1}^{(l)}$, the height of the $i$th column for the $l$th Young tableaux, and $n_l$ the number of columns. In the canonical partition function $\ZC(N)$, $N$ is the number of instantons, i.e. the total number of boxes,
\begin{equation}
\sum_{l=1}^{N_c}\sum_{i=1}^{n_l}\l_i^{(l)}=N.
\end{equation}

In the NS limit, the grand-canonical free energy is roughly equal to the canonical one at large $N$, with $N\e=\g$ fixed (see \ref{rel_FE}). It leads to consider Young tableaux with infinitely many boxes. It is then argued that the summation is dominated by a certain Young tableaux profile determined by extremizing the summation. This profile is characterized by a shape function $f(x)$ \cite{Nekrasov2003a}, or equivalently by a density of instantons defined as
\begin{equation}
\brho\inst(x)=\dfrac{\e_1\e_2}{\e_1+\e_2}\sum_{l=1}^{N_c}\sum_{(i,j)\in\l_i^{(l)}}\d(x-\phi_{l,i,j}).
\end{equation}
We have shown in \ref{rel_densities} that the Bethe roots density $\rho_B(x)$ is minus the derivative of the instanton density $\rho(x)$. In the NS limit, the instanton density is the same for canonical and grand-canonical models, and can be identified with $\brho\inst(x)$. It was further shown in \cite{Bourgine2012a} that the derivative of $\brho\inst(x)$ is a difference of two densities:
\begin{equation}
\rho_B(x)\simeq-\dfrac{d}{dx}\brho\inst(x)\simeq \rho\full(x)-\rho\pert(x).
\end{equation}
In this problem, the numbers of columns $n_l$ is a natural cut-off. It can be sent to infinity, assuming that $\l_i^{(l)}$ is vanishing for large $i$. Then, the two densities in the RHS of the previous equation are formally given by
\begin{equation}
\rho\pert(x)=\sum_{l=1}^{N_c}\sum_{i=1}^{\infty}\d(x-t_{l,i}^0),\quad \rho\full(x)=\sum_{l=1}^{N_c}\sum_{i=1}^{\infty}\d(x-t_{l,i}),
\end{equation}
with $t_{l,i}^0=a_l+(i-1)\e_1$ and $t_{l,i}=t_{l,i}^0+\l_i^{(l)}\e_2$. The density $\rho\pert(x)$ is associated to the perturbative contribution to the gauge theory partition function. It satisfies
\begin{equation}
\rho\pert(x)-\rho\pert(x-\e_1)=\sum_{l=1}^{N_c}\d(x-a_l).
\end{equation}
The relation between densities leads to express ratios of Q-functions as
\begin{equation}\label{qpsi}
\dfrac{q(x)}{q(x-\e_1)}=\dfrac{\psi(x)}{A(x)\psi(x-\e_1)},\quad\text{with}\quad \psi(x)=\prod_{l=1}^{N_c}\prod_{i=1}^\infty(x-t_{l,i}).
\end{equation}
A priori, the infinite product in the definition of $\psi(x)$ requires regularization. However, this function only appears here in well-defined ratios.

The Baxter TQ relation associated to the Bethe roots system of section \refOld{sec_Bethe} is obtained by introducing a function $t(x)$ such that
\begin{equation}
t(x)q(x)=q(x+\e_1)-qQ(x)q(x-\e_1).
\end{equation}
Introducing the relation \ref{qpsi}, and denoting $P(x)=t(x)A(x+\e_1)$, we recover the TQ equation established in \cite{Fucito2011},
\begin{equation}\label{qSW}
P(x)\psi(x)=\psi(x+\e_1)-qM(x)\psi(x-\e_1).
\end{equation}
Comparing this TQ equation with the previous one, we observe that the potential $Q(x)$ is replaced by $M(x)$. As explained in \cite{Bourgine2012a}, this is an effect of the perturbative term: the gauge polynomial dependence in $Q(x)$ cancels with the cross-term between $\rho\full$ and $\rho\pert$. We conclude that, up to the perturbative term $A(x)$, Bethe roots correspond to the shifted height of Young tableaux columns $t_{l,i}$. The equation \ref{qSW} is interpreted as a non-perturbative (or quantized) version of the Seiberg-Witten curve \cite{Poghossian2010,Fucito2011,Mironov2010a,Mironov2010,Popolitov2010,Zenkevich2011,Mironov2012a}.

\paragraph{Confinement} Confinement can also be understood using the Young tableaux representation, where boxes play a role similar to the vertices in clusters. Consider a box $x$ in a Young tableau $\l^{(l)}$ that will be the analogue of the clusters' root. Boxes of the same columns correspond to instantons at a microscopic distance $\sim\e_2$ of $x$, like the fixed vertices with respect to the root of a cluster, or quarks of the same hadron. Other boxes describe instantons at a macroscopic distance $\geq \e_1$ of $x$, like the quarks of other hadrons. For $N_c=1$, there is only one Young tableau and the analogy is actually exact: quarks correspond to boxes, and hadrons to columns where the height gives the number of constituents.  For a general $SU(N_c)$ gauge group, the picture is a bit more complicated because eigenvalues are scattered between several Young tableaux, but a similar sketch can be drawn.

\section{Discussion}
In this paper, we presented a derivation of the TBA-like equation for the Nekrasov instanton partition functions of $\mathcal{N}=2$ theories in the NS limit. It is based on the Mayer cluster expansion, and makes use of generating functions for rooted clusters. The main result is the expression of the free energy at leading order as an on-shell action \ref{action}. The method is applied to a larger class of models with $\a\neq1$, although the corresponding function $L_\a(x)$ remains to be computed. It would be interesting to know if this one parameter generalization of the TBA equation is relevant to integrable problems. The extension to quiver gauge theories has not been considered here, but should be rather straightforward. In this case, different types of vertices should be introduced, being associated to each gauge groups. The generating functions of rooted clusters now wear an index to distinguish the types of roots. Similar graphical relations can be established among them, thus generalizing the equations of motion obtained here.

The phenomenon behind the simplification of the summation over clusters in the NS limit is a confinement of eigenvalues, as shown by the study of the canonical ensemble in the second section. The limit $\a\to0$ was also investigated, and exhibits agreement with the results from collective field theory. Finally, connections with integrable systems was also briefly discussed, as well as the alternative approach based on summations over Young tableaux.

This paper only initiates the study of models defined by \ref{def_CZ}, and there is a lot more to be understood. In particular, the connection with matrix model techniques remains mysterious. Those were investigated in \cite{Bourgine2013} for the case $\a=0$, but these results do not extend easily to the general case. New techniques must be developed to handle confinement in loop equations. In this perspective, the derivation of the subleading order in $\e$ may be a very profitable exercise. At the moment, we are still far from generalizing the topological recursion developed in \cite{Eynard2007a,Borot2013} to models like \ref{def_CZ}. Such a powerful tool would shed a new light on the AGT correspondence: on the Liouville side, correlators takes the form of a \bens\ partition function to which a (quantized) topological recursion applies.

The method developed here may also be useful in understanding the underlying algebraic structures. It was shown in \cite{Schiffmann2012,Kanno2013} (see also \cite{Kanno2012,Kanno2011}) that a central spherical Hecke algebra, SHc, acts on the instanton partition functions. It is natural to expect that this Hopf algebra reduces to the more familiar Yangian structures of integrable systems as $\e_2\to0$. The method presented here to handle the NS limit may help to understand the reduction of symmetry algebra.

\section*{Acknowledgements}
I would like to thank Davide Fioravanti, Yutaka Matsuo, Daniel Ricci-Pacifici and Dima Volin for valuable discussions and comments. I acknowledge the Korea Ministry of Education, Science and Technology (MEST) for the support of the Young Scientist Training Program at the Asia Pacific Center for Theoretical Physics (APCTP).

\appendix
\section{Factorization of the potential dependence for rooted clusters with p-links}\label{AppA}
In this appendix, we provide some arguments for the factorization of the potential leading to the formula \ref{factor_Yp}. The clusters' symmetry coefficients play no role in the discussion, and we focus only on the integral contributions. At this level, rooting a vertex $x$ is equivalent to fix the corresponding integration variable $\phi_x\equiv x$. We will assume that the potential $Q(x)$ can be decomposed into a sum over single poles in the upper half plane (minus infinity) denoted $q_r$, with residues $Q_r$, and a regular part $\Qreg(x)$ as
\begin{equation}\label{decompose_Q}
Q(x)=\Qsing(x)+\Qreg(x),\quad \Qsing(x)=\sum_r{\dfrac{Q_r}{x-q_r}},
\end{equation}
where $\Qreg(x)$ has no singularities inside the contour of integration. For simplicity, we also set $q=1$, as it can be re-absorbed within $Q(x)$. This form of the potential is relevant to the case of Nekrasov partition functions. Later, it will be necessary to replace the potential by a dressing function, in which case double poles may appear. We suggest to treat the double poles at $x=q_r$ with $0<\Im q_r<\infty$ by shifting upward $q_r$ of a distance $k\e$ ($k>0$) in one of the factors. Formally,
\begin{equation}
\int{\dfrac{f(x)}{(x-q_r)^2}\dfrac{dx}{2i\pi}}\simeq \int{\dfrac{f(x)}{(x-q_r)(x-q_r-k\e)}\dfrac{dx}{2i\pi}}=\dfrac{f(q_r+k\e)-f(q_r)}{k\e}\simeq f'(q_r).
\end{equation}
A similar treatment should be performed for higher order poles. One of the subtle points in the present demonstration is to show that this manipulation is perfectly valid at first order in $\e$, and does not influence the final result. This is not guarantee a priori because of the strong sensibility of the kernel $p$ in the precise position of the poles, in particular within p-cycles. We shall come back to this point below.

\subsection{Trees}\label{A1}
We first restrict ourselves to the case of rooted cluster $\D_l^x$ with a tree structure. As a warm-up, consider the simplest case of $l=2$ vertices, i.e. a root attached to a single leaf. The corresponding integral writes
\begin{equation}\label{I-single}
I(x)=Q(x)\int{\dfrac{Q(\phi)d\phi}{2i\pi}\dfrac{\a\e^2}{(x-\phi)^2-\e^2}}.
\end{equation}
To evaluate the integral over $\phi$, we use the decomposition \ref{decompose_Q} of $Q(\phi)$, leading to the sum over residues
\begin{equation}
\dfrac{I(x)}{Q(x)}=\hf\a\e\Qreg(x+\e)+\a\e^2\sum_rQ_r\left[\dfrac1{(x-q_r)^2-\e^2}+\dfrac1{2\e}\dfrac1{x+\e-q_r}\right].
\end{equation}
From the identity
\begin{equation}
\dfrac1{(x-q_r)^2-\e^2}+\dfrac1{2\e}\dfrac1{x+\e-q_r}=\dfrac1{2\e}\dfrac1{x-\e-q_r},
\end{equation}
we can reform $\Qsing$ and write
\begin{equation}
I(x)=\hf\a\e Q(x)Q^+(x),\quad Q^+(x)=\Qreg(x+\e)+\Qsing(x-\e).
\end{equation}
We notice that the poles of $\Qsing(x)$ that would contribute to an integration over $x$ are shifted upward in $Q^+$, and would still contribute. In a similar way, possible singularities of $\Qreg(x)$ are shifted downward, and will not contribute to an integration. Then, we can safely replace $Q^+(x)\simeq Q(x)$ when integrating over $x$. Neglecting the poles of $Q(\phi)$ in the integration performed previously would have led to the opposite shift of $\Qsing$, namely $\Qsing(x+\e)$ which would have been ambiguous for later $x$-integration. Thus, those poles are necessary in the intermediate steps, and cannot be neglected. Only at the final stage, it is possible to write
\begin{equation}
I(x)\simeq\hf\a\e Q(x)^2=Q(x)^2\int{\dfrac{d\phi}{2i\pi}\dfrac{\a\e^2}{(x-\phi)^2-\e^2}}.
\end{equation}
As a side remark, let us also mention that for $Q$ given by \ref{decompose_Q}, we can further compute the contribution of a non-rooted cluster with two vertices,
\begin{equation}
I=\int{I(x)\dfrac{dx}{2i\pi}}=\a\e\sum_r{Q_r\Qreg(q_r)}.
\end{equation}
It is worth noticing here that the squared terms $\Qsing(x)^2$ and $\Qreg(x)^2$ have vanishing contribution, and only the cross-term remains. Since by definition $\Qreg(x)$ is regular at $x=q_r$, everything is well-defined.

\begin{figure}[!t]
\centering
\includegraphics[width=8cm]{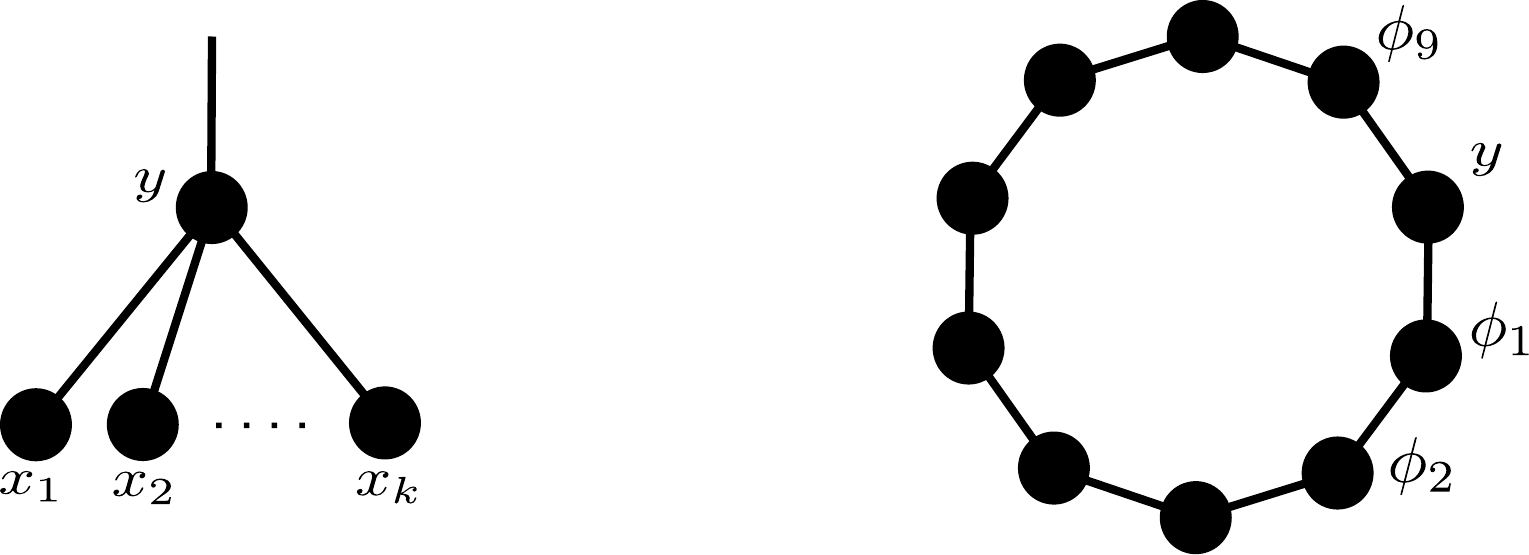}
\caption{Left: Detail of a tree showing some of its deepest leaves. Right: Necklace diagram with $l=10$ vertices.}
\label{dtree}
\end{figure}

As a next step, we should explain how the previous factorization property propagates to any rooted tree by recursion. To do so, let us consider a rooted tree $\D_l^x$, and pick up one of the deepest leaf. We remind the reader that a leaf is a vertex attached to only a single other vertex, and the depth is the minimal distance between the vertex and the root, i.e. the minimal number of intermediate vertices plus one. This leaf is attached to another vertex, that we denote $y$, with a depth $d_\text{max}-1$. The vertex $y$ is attached to a number of leaves $x_1,\cdots,x_k$ (including the previous one), of maximal depth $d_\text{max}$, and a single other vertex of depth $d_\text{max}-2$ (see the configuration on the figure \refOld{dtree}). The corresponding integral is
\begin{equation}
I(x)=\int{\dfrac{dy}{2i\pi}\int{\prod_{i=1}^k\dfrac{Q(x_i)dx_i}{2i\pi}\dfrac{\a\e^2}{(y-x_i)^2-\e^2}J(x,y)}},
\end{equation}
where $J(x,y)$ denotes the contribution of the bi-rooted tree obtained from $\D_l^x$ by removing the leaves $x_1,\cdots x_k$ and rooting the vertex $y$. Integrals over $x_i$ are decoupled, and correspond to $k$ copies of the integral \ref{I-single}. Performing the integrations, we get
\begin{equation}\label{mpoles}
I(x)=\left(\dfrac{\a\e}2\right)^k\int{\dfrac{dy}{2i\pi}Q^+(y)^kJ(x,y)}\equiv \int{\dfrac{Q^+(y)^kdy}{2i\pi}\int{\prod_{i=1}^k\dfrac{dx_i}{2i\pi}\dfrac{\a\e^2}{(y-x_i)^2-\e^2}J(x,y)}}.
\end{equation}
Thus, the potential of the leaves can be transposed to the vertex $y$, at the price of small shifts in $\e$. We have checked before that those shifts are not harmful since no pole would cross the integration contour, and it is safe to replace $Q^+(y)\simeq Q(y)$ (although this replacement will be done only at the end of the computation).  We may assume, after proper treatment of the multiple poles, that the new potential $Q(y)Q^+(y)^k$ associated to the vertex $y$ has still the form \ref{decompose_Q}. Then, the operation of integration over the deepest leaves can be repeated until all vertices are removed, and only the root remains. At the end of the process, the potential appears at the root, modulo $\e$-shifts that can be neglected at first order,
\begin{equation}\label{ptrees}
I(x)\simeq\left(\dfrac{\a\e}2\right)^{l-1}Q(x)^l\equiv Q(x)^l\int{\prod_{\superp{i\in V(\D_l^x)}{i\neq x}}\dfrac{d\phi_i}{2i\pi}\prod_{<ij>\in E(\D_l^x)}\dfrac{\a\e^2}{\phi_{ij}^2-\e^2}}.
\end{equation}
Thus, we have shown that the potential factorizes out of the integrations. A slightly more general result can be established, allowing a different potential $Q_i(\phi_i)$ for each vertex $i$,
\begin{equation}\label{P1}
Q_x(x)\int{\prod_{\superp{i\in V(\D_l^x)}{i\neq x}}\dfrac{Q_i(\phi_i)d\phi_i}{2i\pi}\prod_{<ij>\in E(\D_l^x)}\dfrac{\a\e^2}{\phi_{ij}^2-\e^2}}\simeq\prod_{i\in V(\D_l^x)}Q_i(x)\int{\prod_{\superp{i\in V(\D_l^x)}{i\neq x}}\dfrac{d\phi_i}{2i\pi}\prod_{<ij>\in E(\D_l^x)}\dfrac{\a\e^2}{\phi_{ij}^2-\e^2}},
\end{equation}
at first order. As a corollary, the result of the integration is independent of the exact distribution of the potentials among the vertices. It implies that the potential can be concentrated on any vertex of the tree. This stronger result will be necessary in the next subsection where the presence of cycles is investigated.

The possibility to consider a different potential at each vertex is a key point to justify the manipulation used to remove the multiple poles that appeared in \ref{mpoles}. Indeed, it is possible to tune the poles $q_r$ with upward $\e$-shifts at each vertex such that double poles never happen in the computation, at each order of the recursion. For such a distribution of potentials, the property \ref{P1} holds. Taking the limit $\e\to0$ as a final step, these infinitesimal shifts drop and we conclude that \ref{P1} holds for any distribution of potentials, with multiple poles at initial or intermediate steps.

\subsection{Cycles}
To extend the result of the previous subsection to any rooted cluster $\D_l^x$, we need to investigate the effect of cycles. To begin with, we consider the necklace, a non-rooted cluster of $l$ vertices and $l$ links, each vertex forming two bonds (figure \refOld{dtree}). We will focus on a specific vertex $y$ and denote the other integration variables $\phi_1,\cdots \phi_{l-1}$. The associated integration is
\begin{equation}
I=\int{\dfrac{dy}{2i\pi}Q(y)\int{\dfrac{\a\e^2}{(\phi_1-y)^2-\e^2}\prod_{i=1}^{l-1}\dfrac{\a\e^2}{(\phi_i-\phi_{i+1})^2-\e^2}\dfrac{Q(\phi_i)d\phi_i}{2i\pi}}},\quad \phi_l\equiv y.
\end{equation}
The integrand can be identified with the contribution of a rooted chain of $l$ vertices, with root $y$, and a deformed potential for $\phi_1$:
\begin{equation}\label{deformed_pot}
Q(\phi_1)\to Q'(\phi_1)=Q(\phi_1)\dfrac{\a\e^2}{(\phi_1-y)^2-\e^2}.
\end{equation}
Considering $y$ as a fixed variable, this potential for $\phi_1$ is perfectly valid, and the pole at $\phi_1=y+\e$ simply plays the role of an additional variable $q_r$ in the decomposition \ref{decompose_Q}. There is however an important difference with the initial potential, which is that shifts of $\e$ become meaningful when integrating over $y$ and cannot be simplified. However, since we are only willing to move the $Q(\phi_1)$ part of the $\phi_1$-potential, we can still apply our result \ref{P1} to this rooted chain to get\footnote{To obtain this result more rigorously, consider the rooted chain of $l$ vertex with deformed potential \ref{deformed_pot} for $\phi_1$. Integrating over $\phi_1$, we obtain the potential $Q'^+(\phi_2)Q(\phi_2)$ for $\phi_2$. Integrating successively over $\phi_2,\phi_2,\cdots,\phi_{l-1}$, we obtain a potential for $y$ containing various shifts of each vertex potential. At this stage, it is not possible to simplify this set of shifts because of the presence of the additional factor in \ref{deformed_pot}. However, we may perform a similar operation for the RHS of \ref{pot_in_y}. Simplification of $\e$-shifts is possible in all the $Q$ factors, leading to the equality \ref{pot_in_y} at first order in $\e$.}
\begin{equation}\label{pot_in_y}
I\simeq\int{\dfrac{dy}{2i\pi}Q(y)^l\int{\dfrac{\a\e^2}{(\phi_1-y)^2-\e^2}\prod_{i=1}^{l-1}\dfrac{\a\e^2}{(\phi_i-\phi_{i+1})^2-\e^2}\dfrac{d\phi_i}{2i\pi}}}.
\end{equation}
As in the case of rooted trees, the issue of multiple poles is resolved by the possibility of choosing a different potential at each vertex, thus allowing infinitesimal tuning of the poles.

We observe again in \ref{pot_in_y} that, at first order in $\e$, the potential can be concentrated on any vertex of the necklace. It is immediate to generalize this result to any (non-rooted) cluster with only one cycle. There always exists a vertex $y$ such that if removed, the cycle is broken and the cluster becomes a tree. As before, the integral associated to the initial cluster can be written as the integral over $dy/2i\pi$ of a rooted tree with a deformed potential \ref{deformed_pot} for one of the vertices attached to $y$. Using the property \ref{P1} for this tree, we show that the potential can be concentrated on $y$, as in \ref{pot_in_y}. This remains true for any distribution of the potentials over the vertices. Thus, the potential can be concentrated on any vertex of the cluster. More precisely for 1-cycle clusters $\D_l$ and any vertex $y\in\D_l$, we have at first order
\begin{equation}\label{P2}
\int{\prod_{i\in V(\D_l)}\dfrac{Q_i(\phi_i)d\phi_i}{2i\pi}\prod_{<ij>\in E(\D_l)}\dfrac{\a\e^2}{\phi_{ij}^2-\e^2}}\simeq\int{\dfrac{dy}{2i\pi}\prod_{i\in V(\D_l)}Q_i(y)\int{\prod_{\superp{i\in V(\D_l)}{i\neq y}}\dfrac{d\phi_i}{2i\pi}\prod_{<ij>\in E(\D_l)}\dfrac{\a\e^2}{\phi_{ij}^2-\e^2}}}.
\end{equation}
As a corollary, the potential can be distributed arbitrarily on the vertices of the cluster.

In order to transfer this results to rooted cluster, we may perform a functional derivation of the previous formula with respect to $Q_x(x)$ where $x$ is the root. Concentrating the potential in $x$ before performing the derivation, we obtain the equivalent of \ref{P1} for rooted clusters with a single cycle. Clusters with several 1-cycles can be treated recursively, choosing the integration variable $y$ such that removing it leads to a rooted cluster with one cycle less to which we may apply \ref{P1}. For arbitrary clusters, it may happens that $y$ is attached to more than two other vertices, but the argument extends smoothly, and any cluster can be reached by the recursion. We conclude that \ref{P1} is true for any rooted cluster $\D_l^x$ with only p-links.

\subsection{G-links}\label{A3}
In order to apply the previous results \ref{P1} and \ref{P2} to \ref{Ap_exp}, we need to replace $Q(x)$ with $Y_G(x)$. This replacement is possible if $Y_G(x)$ obeys the property \ref{decompose_Q}, which depends on the exact form of the kernel $G(x)$. In the case of an even kernel with only poles singularities in the upper half-plane, such as those given in \ref{def_G}, the replacement is possible. Let us for instance consider a rooted cluster integral $I(y)$ and attach a G-link $<xy>$ to $y$, such that the new root is $x$. Assume that $I(y)$ satisfies \ref{decompose_Q}, and take the kernel relevant to $\mathcal{N}=2$ SYM in \ref{def_G},\footnote{More involved kernel can be obtained by linear combination of different values of $\e_1$.} one obtains
\begin{equation}\label{int_G}
\int{\dfrac{dy}{2i\pi}G(x-y)I(y)}=-I_\text{reg.}(x+\e_1)-I_\text{sing.}(x-\e_1),
\end{equation}
i.e. that poles of $I(y)$ in the upper half-plane are moved upward and those of the lower half-plane downward, leaving the result well-defined, and obeying again the relation \ref{decompose_Q}. Attaching several vertices by G-links may bring multiple poles, as in \ref{mpoles}, but these can be regularize in a similar way. Since $Y_G(x)$ is a sum over an infinite number of clusters, it may however happen that $Y_G(x)$ does not obey the property \ref{decompose_Q} although we have the invariance of \ref{decompose_Q} under convolution with $G$, as in \ref{int_G}. This issue can be treated by recursion on the number of vertices, which provides a natural cut-off. From \ref{eom_YG} it is seen that each step simply involves a G-convolution. Finally, let us briefly mention that the possibility to replace $Q(x)$ with contributions of $G$-trees in the section \ref{A1} allows to show that mixed trees of $l$ vertices are also of order $O(\e^{l-1})$.

\section{Evaluation of the integral $I_l$}\label{AppB}
\subsection{Leading order in $\a$}
To evaluate $I_l$ at leading order in $\a$, it is better to consider the expression \ref{factor_Yp} of $c_l$. At first order, only trees contribute. We have already shown by recursion that these integrals give a contribution independent of the trees exact structure. This is the result \ref{ptrees} with $Q(x)=1$. It remains to compute the sum over symmetry factors, which can be done using a formula derived in \cite{Bourgine2013},
\begin{equation}
\sum_{\D_l^x}\dfrac1{\s(\D_l^x)}=l\sum_{\D_l}\dfrac1{\s(\D_l)}=\dfrac{l}{l!}\sum_{\D_l}n(\D_l)=\dfrac{l^{l-1}}{l!}
\end{equation}
where the last equality uses the Cayley formula for the number of labeled trees. Putting everything together, we find \ref{Il_1st} for $I_l$ at first order in $\a$.

\subsection{Special case $\a=1$}
To evaluate the integrals $I_l$ in the case $\a=1$, we use a method similar to what was done in \cite{Moore1997}. The Cauchy determinant formula
\begin{equation}
(-1)^l\e^{-l}\prod_{i\neq j}\dfrac{\phi_{ij}}{\phi_{ij}-\e}=\sum_{\s\in\S_l}{(-1)^\s\prod_{i=1}^l\dfrac1{\phi_i-\phi_{\s(i)}-\e}},
\end{equation}
allows to expand $I_l$ on the permutations of the symmetric group $\S_l$,
\begin{equation}
I_l=\dfrac{(-1)^l\e}{(l-1)!}\sum_{\s\in\S_l}{(-1)^\s\int{\prod_{i=1}^{l-1}\dfrac{d\phi_i}{2i\pi}\prod_{i=1}^l\dfrac1{\phi_i-\phi_{\s(i)}-\e}}}.
\end{equation}
These integrals are non-vanishing only if the permutation is a cycle of maximal length $l$. Indeed, permutations can be decomposed into cycles and the integral factorizes into cycle contributions. Cycles of length $m$ that do not contain the fixed variable $\phi_l=x$ give, after re-labeling the variables, the contribution ($\phi_{m+1}\equiv\phi_1$)
\begin{equation}
\int{\prod_{i=1}^{m}\dfrac{d\phi_i}{2i\pi}\prod_{i=1}^m\dfrac1{\phi_i-\phi_{i+1}-\e}}=\int{\dfrac{d\phi_m}{2i\pi}\dfrac1{\phi_m-(\phi_m+(m-1)\e)-\e}},
\end{equation}
which is vanishing, according to \ref{zero_volume}. Thus, only cycles of maximal length $l$ remain. There are $(l-1)!$ such cycles, and their signature is $(-1)^{l-1}$. By re-labeling the variables, we easily show that all these cycles give the same contribution, which can be evaluated by choosing $\s(i)=i+1$ modulo $l$. Residues give $-1/(l\e)$, and
\begin{equation}
I_l=\dfrac{(-1)^l\e}{(l-1)!}\times(l-1)!\times(-1)^{l-1}\times -\dfrac1{l\e}=\dfrac1l.
\end{equation}

\section{Extension of previous results}\label{AppC}
The paper \cite{Bourgine2013} presents several results obtained for our model \ref{def_CZ} in the simplifying limit $\a=0$. At present, it is not known how to extend the matrix model techniques studied there to $\a\neq0$. Nevertheless, many results obtained in \cite{Bourgine2013} does not require a strong assumption on the form of the kernel $f$, and easily generalize to our case. They are summarized in this appendix.

\subsection{Action of $q\p_q$ and densities}
The action of $q\p_q$ on the grand-canonical free energy $\FGC=\e\log\ZGC$ is a purely combinatorial result that applies for any kernel and potential,
\begin{equation}
q\p_q\FGC=\int{Y(x)\dfrac{dx}{2i\pi}}.
\end{equation}
Introducing generating function of $n$-rooted clusters, it can be generalized into
\begin{equation}
q\p_qY(x_1,\cdots,x_n)=\int{Y(x_1,\cdots,x_n,y)\dfrac{dy}{2i\pi}}+nY(x_1,\cdots,x_n),\quad q^n\p_q^n\FGC=\int{Y(x_1,\cdots,x_n)\prod_{i=1}^n\dfrac{dx_i}{2i\pi}}.
\end{equation}

The generating functions $Y$ are related to the grand-canonical densities. Introducing the grand canonical vev of an operator $\Op(x)$,
\begin{equation}
\la\Op(x)\ra=\dfrac1{\ZGC(q)}\sum_{N=0}^\infty\dfrac{q^N\e^{-N}}{N!}\ZC(N)\laN\Op(x)\raN,
\end{equation}
in terms of the canonical vevs,
\begin{equation}
\laN\Op(x)\raN=\dfrac1{\ZC(N)}\int_{\mathbb{R}^N}{\Op(x)\prod_{i=1}^NQ(\phi_i)\dfrac{d\phi_i}{2i\pi}\prod_{\superp{i,j=1}{i<j}}^N\left(1+\e f(\phi_i-\phi_j)\right)},
\end{equation}
the one-point grand-canonical density is defined as
\begin{equation}\label{def_rGC}
\rho_\text{GC}(x)=\e\la\CD(x)\ra,\quad \CD(x)=\sum_i\d(x-\phi_i).
\end{equation}
It is simply related to $Y(x)$ as $2i\pi\rho_\text{GC}(x)=Y(x)$. A similar result holds for the 2-points density,
\begin{equation}\label{brho_Y_2pts}
\rho_\text{GC}(x,y)=\e\la\CD(x)\CD(y)\ra_c=\dfrac1{(2i\pi)^2}Y(x,y)+\dfrac1{2i\pi}\d(x-y) Y(x).
\end{equation}
The relations of this subsection are valid at all order in $\e$.

\subsection{Relation with the canonical partition function at large $N$}
The relations between grand-canonical and canonical ensemble derived in \cite{Bourgine2013} are obtained from the inversion \ref{invert_Laplace} of the discrete Laplace transform. As such, they are not model dependent and still valid in our case. Thus, at large $N$ the canonical free energy is related to the grand-canonical one through the Legendre transform
\begin{equation}
N\left[\FC(N,\e)+1-\log(N\e)\right]\simeq\frac1\e\FGC(q,\e)-N\log q,\quad\text{when}\quad q\p_q\FGC(q,\e)=N\e,
\end{equation}
where free energies are defined as
\begin{equation}
\FGC(q,\e)=\e\log\ZGC(q,\e),\quad\FC(N,\e)=\frac1N\log\ZC(N,\e).
\end{equation}
In the scaling limit $\e\to0,\ N\to\infty$ with $\e N=\g$ fixed, this relation becomes at first order
\begin{equation}\label{rel_FE}
\FC^{(0)}(\g)+1-\log(\g)=\frac1\g\FGC^{0}(q)-\log q,\quad\text{when}\quad q\p_q\FGC^{(0)}(q)=\g,
\end{equation}
where $\log q$ and $\g$ are conjugated variables, and
\begin{equation}
\FC^{(0)}(\g)=\lim_{N\to\infty}\frac1N\log\ZC(N,\g/N).
\end{equation}
A similar relation also applies to densities at leading order in $\e$: at one-point,
\begin{equation}\label{rel_rho}
\rho_\text{C}(x)=\frac1N\laN\CD(x)\raN\implies \rho_\text{GC}(x)\simeq\g\rho_\text{C}(x),
\end{equation}
and at two-points,
\begin{equation}\label{rel_2pts}
\g\rho_\text{C}(x,y)\simeq\rho_\text{GC}(x,y)-\dfrac1{n}\int{\rho_\text{GC}(x,u)du}\int{\rho_\text{GC}(y,v)dv},\quad n=\int{\rho_\text{GC}(x,y)dxdy}.
\end{equation}


\begin{thebibliography}{10}

\bibitem{Alday2009}
L.~Alday, D.~Gaiotto, and Y.~Tachikawa.
\newblock {Liouville Correlation Functions from Four-dimensional Gauge
  Theories}.
\newblock {\em Lett. Math. Phys.}, 91:167--197, 2010.

\bibitem{Wyllard2009}
N.~Wyllard.
\newblock {A\_{N-1} conformal Toda field theory correlation functions from
  conformal N=2 SU(N) quiver gauge theories}.
\newblock {\em JHEP}, 11:002, 2009.

\bibitem{Alba2010}
V.~A. Alba, V.~A. Fateev, A.~V. Litvinov, and G.~M. Tarnopolsky.
\newblock {On combinatorial expansion of the conformal blocks arising from AGT
  conjecture}.
\newblock {\em Lett.Math.Phys.98:33-64,2011}, May 2011.

\bibitem{Fateev2011}
V.~A. Fateev and A.~V. Litvinov.
\newblock {Integrable structure, W-symmetry and AGT relation}.
\newblock {\em JHEP 1201 (2012) 051}, 2011.

\bibitem{Morozov2013}
A.~Morozov and A.~Smirnov.
\newblock {Finalizing the proof of AGT relations with the help of the
  generalized Jack polynomials}, 2013.

\bibitem{Dijkgraaf2009}
R.~Dijkgraaf and C.~Vafa.
\newblock {Toda Theories, Matrix Models, Topological Strings, and N=2 Gauge
  Systems}.
\newblock 2009.

\bibitem{Fujita2009}
M.~Fujita, Y.~Hatsuda, and Ta-Sheng Tai.
\newblock {Genus-one correction to asymptotically free Seiberg-Witten
  prepotential from Dijkgraaf-Vafa matrix model}.
\newblock {\em JHEP 1003:046,2010}, 2009.

\bibitem{Mironov2010c}
A.~Mironov, A.~Morozov, and Sh. Shakirov.
\newblock {Conformal blocks as Dotsenko-Fateev Integral Discriminants}.
\newblock {\em Int. J. Mod. Phys.}, A25:3173--3207, 2010.

\bibitem{Mironov2010e}
A.~Mironov, Al. Morozov, and A.~Morozov.
\newblock {Conformal blocks and generalized Selberg integrals}.
\newblock {\em Nucl. Phys.}, B843:534--557, 2011.

\bibitem{Itoyama2010a}
H.~Itoyama and T.~Oota.
\newblock {Method of Generating q-Expansion Coefficients for Conformal Block
  and N=2 Nekrasov Function by beta-Deformed Matrix Model}.
\newblock {\em Nucl. Phys.}, B838:298--330, 2010.

\bibitem{Itoyama2011}
H.~Itoyama and N.~Yonezawa.
\newblock {$\epsilon$-Corrected Seiberg-Witten Prepotential Obtained From Half
  Genus Expansion in beta-Deformed Matrix Model}.
\newblock {\em Int. J. Mod. Phys.}, A26:3439--3467, 2011.

\bibitem{Nishinaka2011}
T.~Nishinaka and C.~Rim.
\newblock {$\beta$-deformed matrix model and Nekrasov partition function}.
\newblock {\em JHEP}, 02:114, 2012.

\bibitem{Bonelli2011}
G.~Bonelli, K.~Maruyoshi, A.~Tanzini, and F.~Yagi.
\newblock {Generalized matrix models and AGT correspondence at all genera}.
\newblock {\em JHEP}, 1107:055, 2011.

\bibitem{Bonelli2011a}
G.~Bonelli, K.~Maruyoshi, and A.~Tanzini.
\newblock {Quantum Hitchin Systems via beta-deformed Matrix Models}.
\newblock 2011.

\bibitem{Baek2013}
Jong-Hyun Baek.
\newblock {Genus one correction to Seiberg-Witten prepotential from
  $\beta$-deformed matrix model}.
\newblock {\em JHEP 1304:120,2013}, 2013.

\bibitem{Nekrasov2009}
N.~Nekrasov and S.~Shatashvili.
\newblock {Quantization of Integrable Systems and Four Dimensional Gauge
  Theories}.
\newblock 2009.

\bibitem{Ginsparg1993}
P.~Ginsparg and G.~Moore.
\newblock {Lectures on 2D gravity and 2D string theory (TASI 1992)}, 1993.

\bibitem{Jevicki1980}
A.~Jevicki and B.~Sakita.
\newblock {The quantum collective field method and its application to the
  planar limit}.
\newblock {\em Nucl. Phys.}, B165:511, 1980.

\bibitem{Jevicki1981}
A.~Jevicki and B.~Sakita.
\newblock {Collective field approach to the large-N limi: Euclidean field
  theories}.
\newblock {\em Nucl.Phys. B}, 185:89--100, 1981.

\bibitem{Mayer1940}
J.~Mayer and M.~G. Mayer.
\newblock {\em {Statistical Mechanics}}.
\newblock 1940.

\bibitem{Mayer1941}
J.~Mayer and E.~Montroll.
\newblock {Molecular distributions}.
\newblock {\em J. Chem. Phys.}, 9:2--16, 1941.

\bibitem{Bourgine2013}
J.-E. Bourgine.
\newblock {Notes on Mayer Expansions and Matrix Models}, 2013.

\bibitem{Moore1997}
G.~Moore, N.~Nekrasov, and S.~Shatashvili.
\newblock {Integrating Over Higgs Branches}.
\newblock {\em Commun.Math.Phys. 209 (2000) 97-121}, 2006.

\bibitem{Nekrasov2003}
N.~Nekrasov.
\newblock {Seiberg-Witten prepotential from instanton counting}.
\newblock {\em Adv. Theor. Math. Phys.}, 7:831, 2004.

\bibitem{Hoppe1982}
J.~Hoppe.
\newblock {\em {Quantum theory of a massless relativistic surface and a
  two-dimensional bound state problem}}.
\newblock PhD thesis, 1982.

\bibitem{Kazakov1998}
V.~Kazakov, I.~Kostov, and N.~Nekrasov.
\newblock {D-particles, Matrix Integrals and KP hierachy}.
\newblock {\em Nucl.Phys. B557 (1999) 413-442}, 1998.

\bibitem{Hoppe1999}
J.~Hoppe, V.~Kazakov, and I.~K. Kostov.
\newblock {Dimensionally Reduced SYM\_4 as Solvable Matrix Quantum Mechanics}.
\newblock {\em Nucl.Phys. B571 (2000) 479-509}, 2000.

\bibitem{Dyson1962}
F.~J. Dyson.
\newblock {Statistical theory of the energy levels of complex systems. I}.
\newblock {\em J. Math. Phys.}, 3:140--156, 1962.

\bibitem{Meneghelli2013}
C.~Meneghelli and Gang Yang.
\newblock {Mayer-Cluster Expansion of Instanton Partition Functions and
  Thermodynamic Bethe Ansatz}, 2013.

\bibitem{Andersen1977}
H.~Andersen.
\newblock {Cluster Methods in Equilibrium Statistical Mechanics of Fluids}.
\newblock {\em Modern Theoretical Chemistry}, 5:1--45, 1977.

\bibitem{Basso2013}
B.~Basso, A.~Sever, and P.~Vieira.
\newblock {In preparation.}

\bibitem{Yang1968}
C.~N. Yang and C.~P. Yang.
\newblock {Thermodynamics of a one-dimensional system of bosons with repulsive
  delta-function interaction}.
\newblock {\em J. Math. Phys.}, 10:1115--1122, 1969.

\bibitem{Zamolodchikov1990a}
Al.B. Zamolodchikov.
\newblock {Thermodynamic Bethe ansatz in relativistic models: scaling 3-state
  Potts and Lee-Yang Models}.
\newblock {\em Nucl.Phys. B}, 342:695--720, 1990.

\bibitem{Destri1994}
C.~Destri and H.~J. de~Vega.
\newblock {Unified Approach to Thermodynamic Bethe Ansatz and Finite Size
  Corrections for Lattice Models and Field Theories}.
\newblock {\em Nucl. Phys. B}, 438:413--454, 1995.

\bibitem{Gaudin1983}
Michel Gaudin.
\newblock {\em {La fonction d'onde de Bethe}}.
\newblock Collection du CEA, 1983.

\bibitem{Fioravanti1996}
D.~Fioravanti, A.~Mariottini, E.~Quattrini, and F.~Ravanini.
\newblock {Excited State Destri - De Vega Equation for Sine-Gordon and
  Restricted Sine-Gordon Models}.
\newblock {\em Phys.Lett.B390:243-251,1997}, 1996.

\bibitem{Bombardelli2008}
D.~Bombardelli, D.~Fioravanti, and M.~Rossi.
\newblock {Large spin corrections in ${\cal N}=4$ SYM sl(2): still a linear
  integral equation}.
\newblock {\em Nucl.Phys.B810:460-490,2009}, 2008.

\bibitem{Fioravanti2010}
D.~Fioravanti and M.~Rossi.
\newblock {The high spin expansion of twist sector dimensions: the planar N=4
  super Yang-Mills theory}.
\newblock {\em Adv.High Energy Phys.2010:614130,2010}, 2010.

\bibitem{Bourgine2012a}
J.-E. Bourgine.
\newblock {Large N techniques for Nekrasov partition functions and AGT
  conjecture}.
\newblock {\em JHEP 1305 (2013) 047}, 2013.

\bibitem{Poghossian2010}
R.~Poghossian.
\newblock {Deforming SW curve}.
\newblock {\em JHEP}, 04:033, 2011.

\bibitem{Fucito2011}
F.~Fucito, J.~F. Morales, D.~Ricci Pacifici, and R.~Poghossian.
\newblock {Gauge theories on Omega-backgrounds from non commutative
  Seiberg-Witten curves}.
\newblock {\em JHEP}, 05:098, 2011.

\bibitem{Corless1997}
R.~M. Corless, D.~J. Jeffrey, and D.~E. Knuth.
\newblock {A Sequence of Series for the Lambert W Function.}
\newblock In B.~W. Char, P.~S. Wang, and W.~Kuchlin, editors, {\em {ISSAC}},
  pages 197--204. ACM, 1997.

\bibitem{Nekrasov2003a}
N.~Nekrasov and A.~Okounkov.
\newblock {Seiberg-Witten theory and random partitions}.
\newblock 2003.

\bibitem{Seiberg1994}
N.~Seiberg and E.~Witten.
\newblock {Monopole Condensation, And Confinement In N=2 Supersymmetric
  Yang-Mills Theory}.
\newblock {\em Nucl. Phys.}, B426:19--52, 1994.

\bibitem{Seiberg1994a}
N.~Seiberg and E.~Witten.
\newblock {Monopoles, duality and chiral symmetry breaking in N=2
  supersymmetric QCD}.
\newblock {\em Nucl. Phys.}, B431:484--550, 1994.

\bibitem{Ferrari2012a}
F.~Ferrari and M.~Piatek.
\newblock {On a singular Fredholm-type integral equation arising in N=2 super
  Yang-Mills theories}.
\newblock {\em Phys. Lett. B}, 718:1142, 2013.

\bibitem{Klemm2008}
A.~Klemm and P.~Sulkowski.
\newblock {Seiberg-Witten theory and matrix models}.
\newblock {\em Nucl. Phys.}, B819:400--430, 2009.

\bibitem{Sulkowski2009}
P.~Sulkowski.
\newblock {Matrix models for $\beta$-ensembles from Nekrasov partition
  functions}.
\newblock {\em JHEP}, 04:063, 2010.

\bibitem{Sulkowski2009a}
P.~Sulkowski.
\newblock {Matrix models for 2* theories}.
\newblock {\em Phys.Rev.D80:086006,2009}, April 2009.

\bibitem{Nekrasov2012}
N.~Nekrasov and V.~Pestun.
\newblock {Seiberg-Witten geometry of four dimensional N=2 quiver gauge
  theories}.
\newblock 2012.

\bibitem{Fucito2012}
F.~Fucito, J.~F. Morales, and D.~Ricci Pacifici.
\newblock {Deformed Seiberg-Witten Curves for ADE Quivers}, 2013.

\bibitem{Nekrasov2013}
N.~Nekrasov, V.~Pestun, and S.~Shatashvili.
\newblock {Quantum geometry and quiver gauge theories}, 2013.

\bibitem{Mironov2010a}
A.~Mironov and A.~Morozov.
\newblock {Nekrasov Functions from Exact BS Periods: the Case of SU(N)}.
\newblock {\em J.Phys.A}, 43:195401, 2010.

\bibitem{Mironov2010}
A.~Mironov and A.Morozov.
\newblock {Nekrasov Functions and Exact Bohr-Sommerfeld Integrals}.
\newblock {\em JHEP}, 04:040, 2010.

\bibitem{Popolitov2010}
A.~Popolitov.
\newblock {On relation between Nekrasov functions and BS periods in pure SU(N)
  case}.
\newblock 2010.

\bibitem{Zenkevich2011}
Y.~Zenkevich.
\newblock {Nekrasov prepotential with fundamental matter from the quantum spin
  chain}.
\newblock {\em Phys. Lett.}, B701:630--639, 2011.

\bibitem{Mironov2012a}
A.~Mironov, A.~Morozov, Y.~Zenkevich, and A.~Zotov.
\newblock {Spectral Duality in Integrable Systems from AGT Conjecture}.
\newblock {\em JETP Lett.}, 07:45, 2013.

\bibitem{Eynard2007a}
B.~Eynard and N.~Orantin.
\newblock {Invariants of algebraic curves and topological expansion}.
\newblock {\em Commun. Num. Theor. Phys.}, 1:347, 2007.

\bibitem{Borot2013}
G.~Borot, B.~Eynard, and N.~Orantin.
\newblock {Abstract loop equations, topological recursion, and applications},
  2013.

\bibitem{Schiffmann2012}
O.~Schiffmann and E.~Vasserot.
\newblock {Cherednik algebras, W algebras and the equivariant cohomology of the
  moduli space of instantons on $\mathbb{A}^2$}, March 2012.

\bibitem{Kanno2013}
S.~Kanno, Y.~Matsuo, and Hong Zhang.
\newblock {Extended Conformal Symmetry and Recursion Formulae for Nekrasov
  Partition Function}, July 2013.

\bibitem{Kanno2012}
S.~Kanno, Y.~Matsuo, and H.~Zhang.
\newblock {Virasoro constraint for Nekrasov instanton partition function}.
\newblock {\em JHEP}, 10:097, 2012.

\bibitem{Kanno2011}
S.~Kanno, Y.~Matsuo, and S.~Shiba.
\newblock {W(1+infinity) algebra as a symmetry behind AGT relation}.
\newblock {\em Phys. Rev.}, D84:026007, 2011.

\end{thebibliography}
\end{document}